\documentclass[%
 reprint,
 amsmath,amssymb, nofootinbib,
aps,
onecolumn
]{revtex4-2}

\usepackage[table,xcdraw]{xcolor}   
\usepackage{amsmath,amssymb,amsthm, graphicx, mathrsfs, pgf,tikz,marvosym,fontawesome,slashed, mathtools,array,bbm,yfonts, adjustbox, adforn, fourier-orns, relsize,stmaryrd,simplewick,youngtab,ytableau,youngtab, epigraph, url, bclogo,arydshln,multirow,dsfont, lipsum, tensor,feynmf, wrapfig, natbib, enumitem,subfig,verbatim}

\usepackage{physics}
\usepackage[left=20mm,right=20mm,top=35mm,bottom=35mm,columnsep=15pt]{geometry} 
\usepackage{placeins,multirow}
\usepackage[T1]{fontenc}
\usepackage[utf8]{inputenc}
\usepackage[nottoc]{tocbibind}
\usetikzlibrary{arrows}
\usetikzlibrary{positioning}
\usepackage[english]{babel}

\usepackage[uppercase, tiny, center]{titlesec}
\usepackage{cancel}
\usepackage{tikz-cd}
\usepackage{amsmath}
\usepackage{graphicx}
\usepackage{setspace}

\definecolor{r}{cmyk}{1,.50,0,.20} 
\usepackage[colorlinks=true, linkcolor=r, urlcolor=r, linktocpage=true, citecolor=r]{hyperref}

\numberwithin{equation}{section}

\setcitestyle{square,comma}
\usepackage{array,mathtools,amssymb,booktabs}
\newcolumntype{C}{>{$}c<{$}}

\newcommand{\T}{\mathcal{T}}

\newcommand{\Z}{\mathcal{Z}}

\begin{document}

\title{\textbf{A stringy realization of a small and positive cosmological constant in dark bubble cosmology}}

\author{U. Danielsson$^{1}$, O. Henriksson$^{2}$, and D.Panizo$^{1}$}
\affiliation{$^1$ Institutionen f\"{o}r Fysik och Astronomi, Box 803, SE-751 08 Uppsala, Sweden }
\affiliation{$^2$ Department of Physics and Helsinki Institute of Physics P.O. Box 64, FI-00014 University of Helsinki, Finland}

\begin{flushright}
	UUITP - 49/22
	
	\vspace{-0.4 cm}
	
	HIP-2022-29/TH
\end{flushright}

\begin{abstract}
In this paper we construct a stringy embedding of the dark bubble model of an expanding 4D cosmology with the help of branes rotating in extra dimensions. The universe rides a bubble which has nucleated in an unstable higher dimensional background. Our construction is therefore a string theoretical realization of Vilenkin's quantum cosmology. While the cosmological constant vanishes at lowest order, loop corrections predicted by the Weak Gravity Conjecture can induce a phenomenologically viable value. The model predicts the existence of large extra dimensions compatible with observational constraints, and we note a possible connection with the dark dimension proposal.
\end{abstract}

\maketitle

\tableofcontents

\section{Introduction}\label{sec:Introduction}
Finding meta-stable vacua in string theory with a positive cosmological constant (CC) $\Lambda$, remains an unsolved problem \cite{Danielsson:2018ztv, Obied:2018sgi, vanbeest2021lectures}.
The dark bubble model differs in crucial aspects from other proposals, with our four dimensional universe riding an expanding bubble whose embedding naturally induces a positive CC \cite{Banerjee:2018qey, Banerjee:2019fzz}. It makes use of ingredients such as strings \cite{Banerjee:2020wix,Banerjee:2020wov}, branes \cite{Banerjee:2021qei}, extra dimensions and anti-de Sitter (AdS) spaces that commonly surface in string theory. For a quick review, see \cite{Banerjee:2021yrb}.

What distinguishes the dark bubble from other attempts to construct dS space, is that it makes explicit use of the notorious instabilities that seem to be present when supersymmetry is broken in string theory (i.e. any vacua in string theory, if non-supersymmetric, is unstable and will decay via the nucleation of branes, as a consequence of the Weak Gravity Conjecture (WGC) \cite{ooguri2017nonsupersymmetric, freivogel2016vacua,2016mtxDanielsson}). The starting point for the dark bubble is a meta-stable $AdS_5$ space that decays into one with a lower energy through the nucleation of a spherical bubble \cite{Banerjee:2018qey}. Our universe (on the co-dimension one boundary) is riding the bubble as it expands. As explored in \cite{Danielsson:2021tyb}, the nucleation event can be identified with the creation event in Vilenkin quantum cosmology.

A remaining challenge has been to find an explicit embedding of the dark bubble into string theory\footnote{In the context of non-supersymmetric string theory, where supersymmetry is broken or absent already at the string scale, dS from dark bubbles has been studied in \cite{Basile:2020mpt}. For another interesting proposal see \cite{Berglund2021}.} In this paper we present progress on this task. In section \ref{sec:10D Background} we review a simple type IIB string theory construction with a stack of \emph{rotating} D3-branes whose near-horizon geometry takes the form of $AdS_5 \times S^5$ perturbed by angular momentum on the $S^5$ \cite{Cvetic_1999}. These solutions are holographic duals of states of $\mathcal{N}=4$ super-Yang-Mills (SYM), living on an $S^3$, at non-zero chemical potential and temperature. The chemical potential, or angular velocity on the gravitational side, breaks the supersymmetry of the system, even in the extremal limit. From a 5D point of view the metric takes the simple form of an AdS-Reissner-Nordström (RN) black hole\footnote{In  \cite{Koga:2022opd}, AdS$_5$-Kerr black holes were used for brane nucleation.}.

In section \ref{sec: Embedding}, we study the dynamics of a D3-brane in such a background using the probe approximation. The background has previously been shown to be unstable to the nucleation of D3-branes, which tunnel from the black hole horizon, through a potential barrier, to a classical turning point outside the horizon \cite{Henriksson:2019zph}. Emerging at rest, such a brane will then start to expand, and we show how a 4D observer will interpret this as a novel FLRW-cosmology. This provides the first stringy realization of the dark bubble model in supersymmetric string theory. In section \ref{sec:Matching} we perform a comparison between the Hamiltonian obtained in section \ref{sec: Embedding} and the Israel junction conditions \cite{Israel:1966rt} used in the dark bubble model \cite{Danielsson:2022fhd}, confirming that the ten dimensional top-down construction unambiguously matches with that of the five-dimensional one. 

In section \ref{sec:Cosmology}, we further analyse the four dimensional cosmology arising from this top-down construction and how the dynamics of the embedded universe are intrinsically related to higher dimensional features, such as the angular momentum of the D3-brane. As a consequence of supersymmetry in the unperturbed background that we start from, the D3-branes are \emph{critical} to leading order, and thus our scenario has a close to vanishing 4D cosmological constant. However, in section \ref{sec: Physical Implications} we take an additional step, making use of WGC to argue that there will be a small higher order deviation from criticality, inducing a tiny positive cosmological constant. We discuss the possible physical consequences, and point out similarities (and differences) with the dark dimension proposed in \cite{Montero:2022prj}. 

We summarise our results in the final section \ref{sec: Conclusions}, and discuss further work needed to establish the dark bubble as a rigorous implementation of dark energy in string theory.

\section{10D background model review}\label{sec:10D Background}

In this section, we review the string backgrounds that we will use to embed the dark bubble. They consist of black brane solutions of type IIB supergravity, sourced by a stack of D3-branes with angular momentum in the transverse directions. Their near-horizon limit is asymptotically AdS$_5\times S^5$, and after a reduction on the $S^5$ they appear as five-dimensional charged black holes. These solutions were introduced in \cite{Behrndt_1999} starting from a consistent truncation to 5D $N=2$ gauged SUGRA, and were uplifted to 10D in \cite{Cvetic__1999}. In general, they support three independent angular momenta, corresponding to charges of the 5D black holes, but we restrict ourselves to the case where these are identical.

By AdS/CFT, the solutions are dual to states of $\mathcal{N}=4$ super-Yang-Mills with non-zero temperature and R-charge density. With this applications in mind, the solutions have been studied by several authors, see e.g. \cite{Cai:1998ji, Cvetic_1999, Cvetic_3, Chamblin_1999, Yamada:2008em, Henriksson:2019zph}. In particular, it was shown in \cite{Yamada:2008em,Henriksson:2019zph} that at sufficiently large R-charge chemical potential, they are unstable to the emission of the D3-branes sourcing them. This was seen by studying a probe D3-brane in a fixed background solution, and showing that its energy is minimized outside the event horizon. In general, however, there exists a potential barrier between the horizon and the true minimum, so D-branes wanting to lower their energy must do so through tunneling, or \emph{nucleation} \cite{Henriksson:2021zei}. 

In the dual $\mathcal{N}=4$ theory, the position of the D3-branes is dual to the expectation value of scalar fields, and the instability indicates that some scalar degrees of freedom want to condense, thus Higgsing the gauge group. This is analogous to the phenomenon of color superconductivity, which is known to occur in QCD at large baryon chemical potential. However, our focus will be mainly on the gravitational side of the duality. We will now review the necessary features of this set up for our purpose. More information and details can be found in \cite{Henriksson:2019zph}.

Our starting point is the line element describing the geometry of the $10D$ background, given by
\begin{equation}\label{eq:10Dmetric}
	ds_{10}^2= ds_5^2 + L^2 \sum_{i=1}^3 \left\{ d\sigma_i^2 + \sigma_i^2 \left( d\phi_i + L^{-1} A(r) \right)^2 \right\} \ ,
\end{equation}
where $L$ is the $AdS_{5}$ radius, which also sets the scale of the extra dimensions. The $\sigma_i$ can be parametrized with the angles on a two-sphere as
\begin{equation}
	\sigma_1 = \sin\theta \ , \qquad \sigma_2 = \cos\theta \sin\psi \ , \qquad \sigma_3 = \cos\theta \cos\psi \ .
\end{equation}
The 5D asymptotically-AdS metric is given by
\begin{equation}\label{eq:5DmetricQ}
	ds_5^2 = - h(r)^{-2} f(r) dt^2
	+ h(r) \big[ f(r)^{-1} dr^2 + r^2 d\Omega_{3}^2 \big] \ ,
\end{equation}
where $d\Omega_{3}^2$ is the usual unit metric of the 3-sphere. The radial functions $h(r)$ and $f(r)$ are expressed as
\begin{align}
	h(r)&=1+\frac{q^2}{r^2} \ , \\
	f(r) &= 1 - \frac{M}{r^2} + \frac{r^2}{L^2} h(r)^3 \ .
\end{align}

In order to simplify our future expressions, it will be useful to substitute the mass parameter $M$ in favor of the horizon radius $r_H$, defined to be the largest root of $f(r)$:
\begin{equation}
	M = r_H^2\left(1 + \frac{r_H^2}{L^2} h(r_H)^3 \right) \ .
\end{equation}

The one-form $A(r)$ appearing in the 10D metric (\ref{eq:10Dmetric}) acts as a gauge field from the 5D point of view, and takes the form
\begin{equation}\label{eq:gaugepots}
	A = \frac{q}{r_H^2+q^2}\sqrt{ (r_H^2 + q^2) + \frac{r_H^4}{L^2} h(r_H)^3} \left(1-\frac{r_H^2+q^2}{r^2+q^2}\right) dt \ .
\end{equation}
We have used the gauge freedom to add a constant that sets $A(r)$ to zero on the horizon. This is necessary in order for it to have finite norm there.

The rotating D3-branes will source the self dual $F_{5}$ Ramond-Ramond (RR) field strength, which can be written in terms of the radial function $h(r)$ and the one-form $A$ \cite{Henriksson:2019zph}. What we will need is the corresponding 4-form potential $C_{4}$, with $d\:C_{4}=F_{5}$, given by
\begin{equation}
	C_4 = \frac{1}{L}\left[(r^2+q^2)^2 - (r_H^2+q^2)^2\right] dt \wedge \epsilon_3 + L^2 q \sqrt{ r_H^2 + q^2 + \frac{r_H^4}{L^2} h(r_H)^3}\ \sum_i \sigma_i^2\, d\phi_i \wedge \epsilon_3 \ , \label{eq:C4}
\end{equation}
where $\epsilon_3$ is the volume form of the unit 3-sphere. This 4-form will be important in the next section, as it enters into the Wess-Zumino (WZ) part of the brane action, thus influencing the brane's dynamics.

Let us now come back to the metric \ref{eq:5DmetricQ}. For computational convenience, it is useful to perform a change of coordinates, by defining
\begin{equation}
	z^2 = r^2h(r) = q^2 + r^2 .
\end{equation}
This change of coordinates will transform \ref{eq:5DmetricQ} into a patch of $AdS_{5}$-RN,
\begin{align}\label{eq:metricargument}
	\textrm{d} s_5^2 = -g(z)\textrm{d} t^2 + g(z)^{-1}\textrm{d} z^2+z^2\textrm{d}\Omega_3^2, && g(z) = 1 - \frac{2\kappa_5 \mu}{z^2}+\frac{\kappa_5^2 \theta^2}{z^4} + k^2z^2
\end{align}
where we defined 
\begin{align}
	2\kappa_5\mu = M + 2q^2, && \kappa_5^2 \theta^2 = q^2(M+q^2), && k = L^{-1},
\end{align}
with $\kappa_5 = 8\pi G_5$. 
For a small black hole, with $z_H<<L$, we are back in flat space and require $\theta<\mu$ to get a horizon and no naked singularity. On the other hand, if we consider an horizon larger that the $AdS_{5}$ radius, we find that $\theta<<\mu$ is needed.

For computational purposes in the incoming sections, it will be more convenient to express (\ref{eq:metricargument}) in terms of the two horizons of a Reissner-Nordström black hole. With $\left\{z_{h}, z_{H}\right\}$ as the inner and outer horizons, we have $g(z_h)=g(z_H)=0$, and make the Ansatz
\begin{equation}
	g(z) = \frac{k^{2}}{z^{4}}\left(z^{2} + c\right)\:\left(z^{2} -z_{h}^{2}\right)\:\left(z^{2} - z_{H}^{2}\right).
\end{equation}
Matching with (\ref{eq:metricargument}) one obtains
\begin{equation}\label{eq:muthetatohorizons}
	\begin{split}
		\kappa_5 \mu &= \tfrac{1}{2}\left(z_h^{2} + z_H^{2} +k^2\left( z_h^4  + z_h^{2} z_H^{2}  + z_H^4 \right)\right),  \\
		\kappa_5^2 \theta^2 &=z_{h}^{2}\:z_{H}^{2} \:\left(1 + k^{2}\left(z^{2}_{h}+ z_{H}^{2}\right)\right),\\
		c &= z_{h}^{2} + z_{H}^{2} + \tfrac{1}{k^{2}},
	\end{split}
\end{equation}
which simplifies further when the black hole is extremal (i.e $z_{h} = z_{H}$). Equipped with this new set of coordinates and relations, we can embark to study the dynamics of the nucleated branes in this rewritten background. This will be the task of the next section.

\section{Embedding the nucleated brane}\label{sec: Embedding}

\subsection{The brane action}

Let us now consider the motion of the nucleated brane in the ten dimensional background. This calculation follows closely the one in \cite{Henriksson:2019zph}, and we refer the reader there for more details. We start from the action of a (probe) D3-brane, which is the sum of a Dirac-Born-Infeld (DBI) term and a WZ term:
\begin{equation}\label{eq:D3action}
	S_{D3} = -T_3\int d^4\xi \sqrt{-\det P[G]}+T_3\int P[C_4]
\end{equation}
The WZ term with $C_4$ captures the non-gravitational forces on the brane due to fields external to it. $P[\dots]$ denotes the pullback of spacetime fields to the world volume of the brane. Note that the dilaton $\phi$ is constant in the solutions of interest. In the following we denote the brane's embedding functions by capital versions of the spacetime coordinates: $X^\mu(\xi)$. 

We assume the brane to wrap the 3-sphere inside the $AdS_5$-piece of the metric, with a radial $z$-coordinate that depends only on time. Since the background black hole rotates in the azimuthal $\phi_i$-directions on $S_{5}$, we must also allow our probe to move in those directions. Since we set the three angular momenta equal in the background, we assume that the brane rotates with the same angular velocity in all $\phi_i$ directions. As will become evident later on, from the point of view of an observer living in $AdS_5$, the rotation translates into an effective charge under the gauge field $A$. When a brane nucleates, carrying a portion of the charge, it will also effectively lower the charge of the remaining black hole\footnote{This is ignored in the probe approximation, but still essential when we compare with the appropriate limit of the Israel junction conditions across the brane.}. The brane's world volume can be parameterized by the angles of the 3-sphere, as well as a timelike coordinate $\tau$ which we keep arbitrary for now. The embedding is then given by
\begin{equation}
	\begin{gathered}
		\T=\T(\tau) \, , \qquad \Z=\Z(\tau) \, , \qquad \Theta = \theta_0 \, , \qquad \Psi = \psi_0 \, , \qquad \Phi_1 = \Phi_2 = \Phi_3 = \Phi(\tau) \ .
	\end{gathered}
\end{equation}
Here, $\theta_0$ and $\psi_0$ are constants --- in the symmetric case we are interested in, nothing will end up depending on them. Derivatives with respect to $\tau$ will be denoted by a dot. The components of the 10D metric $ds_{10}^2$ will be denoted by $G_{\mu\nu}$, and $(C_4)_t$ and $(C_4)_{\phi_i}$ will denote the coefficients of $dt \wedge \epsilon_3$ and $d\phi_i \wedge \epsilon_3$ in (\ref{eq:C4}), respectively. It will be useful to define $G_{t\phi}=\sum_i G_{t\phi_i}$, $G_{\phi\phi}=\sum_i G_{\phi_i\phi_i}$, as well as $(C_4)_{\phi}=\sum_i(C_4)_{\phi_i}$.

The velocity of a comoving observer on the brane is
\begin{equation}
	\dot X \equiv \frac{\partial X^\mu}{\partial\tau}\partial_\mu = \dot \T \partial_t + \dot \Z\, \partial_z + \dot \Phi\sum_{i=1}^{3} \,\partial_{\phi_i} \ ,
\end{equation}
whose square is
\begin{equation}\label{eq:10velocity}
	\dot X^2 = \dot X_\mu \dot X^\mu = G_{tt} \dot \T^2 + G_{zz} \dot \Z^2 + 2G_{t\phi} \dot \Phi \dot \T + G_{\phi\phi} \dot\Phi^2 \ .
\end{equation}
Then, the induced line element on the brane worldvolume can be written as
\begin{equation}
	ds^2_4 = \dot X^2 d\tau^2 + \Z(\tau)^2 d\Omega_{3}^2 \ .
\end{equation}
We note that picking $\tau$ to be the proper time of a brane-bound observer would set $\dot{X}^2=-1$, and the induced metric then takes the standard FLRW-form with $\Z(\tau)$ acting as the scale factor.

The action for a probe D3-brane finally takes the form
\begin{equation}
	\begin{split}
		S_{D3} &= -T_3 \int d\tau \wedge \epsilon_3 \, \left\{ \Z^3 \sqrt{ -\dot X^2 } - (C_4)_t \dot\T - (C_4)_{\phi} \dot\Phi \right\} \\
		&= -2\pi^2 T_3 \int d\tau \left\{ \Z^3 \sqrt{ -\dot X^2 } - (C_4)_t \dot\T - (C_4)_{\phi} \dot\Phi \right\} \equiv \int d\tau \mathcal{L}_{D3} ,
	\end{split}
\end{equation}
where in the second line we performed the angular integrals over the $S^3$.

\subsection{The probe D-brane Hamiltonian}

It is useful to define the conserved angular momentum:
\begin{equation}
	J \equiv \frac{1}{2\pi^2 T_3}\frac{d\mathcal{L}_{D3}}{d\dot\Phi} = \frac{\Z^3 }{\sqrt{ -\dot X^2 }} \left(G_{t\phi} \dot\T + G_{\phi\phi} \dot\Phi \right) + (C_4)_\phi.
\end{equation}
One can then Legendre transform to substitute $\dot\Phi$ for $J$, automatically satisfying $\Phi$'s EoM. Defining $J_C = J-(C_4)_\phi$, we find
\begin{equation}
	\dot\Phi = -\frac{G_{t\phi}}{G_{\phi\phi}}\dot\T \pm J_C \sqrt{\frac{\left(\frac{G_{t\phi}^2}{G_{\phi\phi}^2}-\frac{G_{tt}}{G_{\phi\phi}} \right) \dot\T^2 - \frac{G_{zz}}{G_{\phi\phi}} \dot\Z^2 }{\left(\Z^6\, G_{\phi\phi} + J_C^2\right)}}.
\end{equation}
and
\begin{equation}
	\begin{split}
		\mathcal{L}_{D3}^J &= \mathcal{L}_{D3} - \dot\Phi J \\
		&= -2\pi^2 T_3 \left\{-(C_4)_t \dot\T - \frac{G_{t\phi}}{G_{\phi\phi}}J_C \dot\T + \sqrt{-\left(\Z^6 + \frac{J_C^2}{G_{\phi\phi}}\right) \left(\left(G_{tt}-\frac{G_{t\phi}^2}{G_{\phi\phi}} \right)\dot\T^2 + G_{zz} \dot\Z^2 \right) } \right\}\\
		&= -2\pi^2 T_3 \dot\T \left\{-(C_4)_t  - \frac{G_{t\phi}}{G_{\phi\phi}}J_C  + \sqrt{-\left(\Z^6 + \frac{J_C^2}{G_{\phi\phi}}\right) \left( G_{tt}-\frac{G_{t\phi}^2}{G_{\phi\phi}} + G_{zz} \Z'^2 \right) } \right\},
	\end{split}
\end{equation}
where now a prime denotes a derivative with respect to the spacetime time-coordinate $t$. This result can be nicely interpreted from a 5D point of view. Denoting the components of the 5D metric $ds_5^2$ by $g_{\mu\nu}$, we can write it as
\begin{equation}\label{eq:LagJ}
	\mathcal{L}_{D3}^J = -2\pi^2 T_3 \dot\T \left\{-(C_4)_t  - \frac{J_C}{L}A_t  + \sqrt{-\left(\Z^6 + \frac{J_C^2}{L^2}\right) \left(g_{tt}+g_{zz} \Z'^2 \right) } \right\} .
\end{equation}
The linear term $J_C$ in (\ref{eq:LagJ}) shows that the brane carries charge and couples to the bulk gauge field. Note that it is $J_C$ rather than $J$ that measures the charge. The relative shift can be understood by frame dragging due to the rotation in the extra dimensions. 
Focusing on the square root term, we note that in the absence of $J_C$, it looks exactly like an effective 5D DBI-term. The modification due to $J_C$ can be interpreted as the addition of a flux that corresponds to a uniform density of particles, or 0-branes, dissolved into the D3-brane. This is how the brane carries the charge from a 5D point of view. To better understand this, we can formally collapse the brane to a point by putting $\Z=0$, and thus find the action $-\frac{2\pi^2 T_3 J_C}{ L}\dot\T  \sqrt{- \left(g_{tt}+g_{zz} \Z'^2 \right)}$. In the absence of the dissolved 0-branes, the total mass of the D3 would approach zero in this limit. Here, however, the mass approaches $\frac{2\pi^2 T_3 J_C}{L}$, and we interpret this as the total mass of the dissolved particles. Lastly, the $(C_4)_t$-term can be interpreted as a 5D WZ-term, since one can define an effective 5D 5-form field $F_5^{eff}=d\big[(C_4)_t dt\wedge\epsilon_3\big]$ with $F_5^{eff}$ proportional to the volume form of $ds_5^2$.

Next, we can further Legendre transform to obtain the Hamiltonian
\begin{equation}\label{eq:Hamiltonian}
	H=2\pi^2 T_3 \dot\T \left( \frac{-g_{tt} \sqrt{\Z^6 + \frac{J_C^2}{L^2}}}{\sqrt{ -\left(g_{tt}+g_{zz} \Z'^2 \right) }} -(C_4)_t  - \frac{J_C}{L}A_t \right),
\end{equation}
given in terms of the proper time of the brane. However, it is important to realize that this is {\it not} the time variable relevant for a 4D observer. Such an observer will be moving together with the nucleated brane in AdS$_5$, but will not be moving together with the brane in the five internal dimensions\footnote{Compare a KK-compactification to 4D. The proper time of a 4D observer, built out of KK-particles, does not involve the internal momenta or velocities of those particles.}. The relevant proper time, $\tilde{\tau}$, is then simply computed as
\begin{equation}
	d\tilde{\tau}
	= dt\sqrt{ -\left(g_{tt}+g_{zz} \Z'^2 \right) }.
\end{equation}
Using
\begin{equation}
	g_{tt}+g_{zz} \Z'^2=  \frac{g_{tt}}{1+g_{zz} \left(\frac{d\Z}{d\tilde{\tau}}\right)^2},
\end{equation}
we then find that the Hamiltonian in terms of the proper time of the 4D observer is given by
\begin{equation} \label{probebranehamiltonian}
	H=2\pi^2 T_3 \frac{d\T}{d \tilde{\tau}} \left( \sqrt{-g_{tt}\left(\Z^6 + \frac{J_C^2}{L^2}\right)}\sqrt{ 1+g_{zz}\left(\frac{d\Z}{d\tilde{\tau}}\right)^2   }-(C_4)_t  - \frac{J_C}{L}A_t\right) .
\end{equation}
Now, we can use AdS-RN coordinates and split the previous Hamiltonian (\ref{probebranehamiltonian}) to find:
\begin{equation} \label{eq: wz}
	(C_4)_t  + \frac{J_C}{L}A_t = \frac{1}{L} \left(\Z^4-z_H^4\right) + \frac{\kappa_{5}\:J_C \:\theta}{L}\left(\frac{1}{z_H^2}-\frac{1}{\Z^2} \right),
\end{equation}
and:
\begin{equation}
	\begin{split}
		\sqrt{-g_{tt}\left(\Z^6 + \frac{J_C^2}{L^2}\right)}\sqrt{ 1+g_{zz} \dot\Z^2  } &=\sqrt{\left(\Z^6 + \frac{J_C^2}{L^2}\right) }\sqrt{ g(\Z )+\dot\Z^2  }\\
		&=\sqrt{\left(\Z^6 + \frac{J_C^2}{L^2}\right) }\sqrt{ k^2 \Z^2+1-\frac{2\kappa_5\mu}{\Z^2}+\frac{\kappa_5 ^2 \theta^2}{\Z^4}+\dot\Z^2}.
	\end{split}
\end{equation}
From now on, we let a dot refer to the proper time $\tilde{\tau}$ of the 4D observer.

As mentioned above, the nucleating D3-brane carries away some charge through the density of dissolved D0-branes. Thus, the remaining black hole will have its charge reduced by the charge of the D3. Hence, the charge in the metric inside of the brane will be shifted accordingly. 

This Hamiltonian should be put equal to a conserved energy $E$, while the constant in $(C_4)_t$ is a gauge choice that also shifts $E$. When we move beyond the probe approximation considering the nucleation event, matching against the junction conditions, we will see how any ambiguity is fixed by energy conservation. 

\section{Matching the Hamiltonian with the junction conditions}
\label{sec:Matching}

The original dark bubble proposals were formulated in five-dimensional AdS space, with the thin-wall bubble represented as a codimension-one shell. This meant that its dynamics could be studied using Israel junction conditions. In this section, we will show that such junction conditions can, in the appropriate limit, reproduce the dynamics described by the probe brane Hamiltonian found in the previous section.

Starting from the equation of motion for the scale factor $\Z$ obtained from the Hamiltonian (\ref{eq:Hamiltonian}), we solve for $\dot{\Z}^{2}$, imposing $H=E=0$ since the bubble will nucleate at rest with, in an approximately extremal background, zero total energy. Then, the EoM controlling the dynamics of the brane is of the form:
\begin{equation}\label{eq:potentialfromH}
	\dot{\Z}^{2} = -f(\Z) -k^2 \Z^2 +  \frac{k^{2}\:\left(\Z^{4}-\Z^{4}_{H}+\frac{\theta \: \kappa_{5}\:J_{c}}{\Z_{H}^{2}}\left(1-\frac{\Z_{H}^2}{\Z^2}\right)\right)^2}{k^{2}\:J_{c}^{2}+\Z^{6}},
\end{equation}
with $f(\Z)=  1 -\frac{2\kappa_5 \mu}{\Z^2}+\frac{\kappa_5^2 \theta^2}{\Z^4}$, such that $g(\Z )= k^2 \Z^2 + f(\Z )$.  The last term corresponds to the contribution of the Wess-Zumino term in the action of the brane. We note that the leading $\Z^2$ term in the large $\Z$-limit cancels, consistent with the tension of the brane equaling the critical one, implying no net 4D cosmological constant. The expression (\ref{eq:potentialfromH}) is simple enough to be compared to the Friedmann equation obtained from the junction condition for the bubble. This will be our next step.

Let us start by consider a thin shell in the form of a brane that tunnels from the horizon $\Z_{H}$, nucleates at radius $\Z > \Z_{H}$ and starts to expand. The shell will be a junction between the outer metric described by $g_+(\Z)$ and the inner metric described by $g_{-}(\Z)$, with
\begin{equation}
	g_{\pm}(\Z)=k_{\pm}^2 \Z^2+1-\frac{2\kappa_5 \mu_{\pm}}{\Z^2}+\frac{\kappa_5^2 \theta_{\pm}^2}{\Z^4},
\end{equation}
The junction condition \cite{Israel:1966rt} for this system becomes:
\begin{equation} \label{eq: junction}
	T_3 = \frac{3}{\kappa_{5}}\left(\sqrt{\frac{g_-(a)}{a^2}+\frac{\dot{a}^2}{a^2}}-\sqrt{\frac{g_+(a)}{a^2}+\frac{\dot{a}^2}{a^2}}\right),
\end{equation}
where $\kappa_5 = 8 \pi G_5$, $\dot{a}$ denotes a derivative with respect to the proper time of a shell observer, and $\Z=a(\tau)$. We can solve exactly the junction conditions for $\dot{a}^{2}$ to obtain the following Friedmann like equation:
\begin{equation}\label{eq: FriedmannfromJC}
	\dot{a}^2= \frac{1}{4 \sigma^{2} a^{2}} \left[\left(k_{-}^{2}-k_{+}^{2}\right)a^{2} +\left(f_{-}(a) - f_{+}(a)\right)\right]^{2} + \frac{\sigma^{2}}{4}a^{2} -\frac{1}{2}\left[\left(f_{-}(a) + f_{+}(a)\right) + \left(k_{-}^{2}+k_{+}^{2}\right)a^{2} \right],
\end{equation}
with $f_{\pm}(a)=  1 -\frac{2\kappa_5 \mu_{\pm}}{a^2}+\frac{\kappa_5^2 \theta_{\pm}^2}{a^4}$ and $\sigma =T_3 \frac{\kappa_5}{3} = (k_- -k_+) = \Delta k$. In order to compare with  (\ref{eq:potentialfromH}), we must make sure to be in the right regime, i.e. that of the probe approximation. This implies that we must work at linear order in the perturbations when making expansions of the form:
\begin{equation}
	k_{\pm}=k\mp \tfrac{1}{2} \Delta k, \quad\quad\quad\quad \theta_{\pm}=\theta\mp \tfrac{1}{2} \Delta \theta, \quad\quad\quad\quad \mu_{\pm}=\mu\mp \tfrac{1}{2} \Delta \mu.
\end{equation}
Note that $\Delta k =k_- - k_+ >0$, while $\Delta \theta= \theta _- - \theta_+<0$ since the black hole is loosing charge due to the nucleation. Substituting these expressions into the previous Friedmann equation, and considering terms up to linear order, we find:
\begin{equation}\label{eq:FriedmannlinearJC}
	\begin{split}
		\dot{a}^2 &= -f(a)+\frac{k}{\Delta k}\left( f_{-}(a)-f_{+}(a)\right)+\frac{1}{4a^2(\Delta k)^2}\left(f_{-}(a)-f_{+}(a)\right)^{2}=\\
		&=-f(a)-k^2 a^2+\frac{k^2}{a^6}\left( a^4 +\frac{a^2}{2k\Delta k}\left( f_{-}(a)-f_{+}(a)\right)\right)^2,
	\end{split}
\end{equation}
such that $f(a)$ carries zeroth order information about the mass $\mu$ and charge $\theta$, while $f_-(a)-f_+(a)$ contains term linear in $\Delta \mu$ and $\Delta\theta$.

Expression (\ref{eq:FriedmannlinearJC}) can be simplified even further. This can be done by evaluating the junction condition and $g(\Z)$ at the horizon $\Z_{H}$. In the probe limit we need both $g_{\pm}(\Z)$ to vanish at the horizon $\Z_{H}$. This implies:
\begin{equation}\label{eq:JCuptolinear}
	\begin{split}
		\textbf{$0^{th}$}&: \quad k^2 a^2 +g(a)=0, \\
		\textbf{$1^{st}$}&: \quad k \Delta k a_H^2-\frac{\kappa_5 \Delta \mu}{a_H^2}+\frac{\kappa_5 \theta \Delta \theta}{ a_H^4}=0.
	\end{split}
\end{equation}
Using (\ref{eq:JCuptolinear}) in expression (\ref{eq:FriedmannlinearJC}) one arrives at:
\begin{equation}\label{eq:FinalfriedmannfromJC}
	\dot{a}^2= -f(a)-k^2 a^2+\frac{k^2}{a^6}\left( a^4-a_H^4 - \frac{\kappa_{5}^{2}\theta \Delta \theta}{a_{H}^{2}\:k \Delta k } \left(1-\frac{a_H^2}{a^2} \right) \right) ^2.
\end{equation}
This should now be compared with what we obtained from the Hamiltonian in eq (\ref{eq:potentialfromH}), with $\Z = a(\tau)$. In fact, the overall structure is identical. The first two terms correspond to the function $g(a)$ in the metric of the $5D$ black hole, while the last term in (\ref{eq:FinalfriedmannfromJC}) requires a deeper analysis.

The first thing one observes is the absence of $J_{c}^2$ in the denominator of the expression (\ref{eq:FinalfriedmannfromJC}), coming from the junction conditions. This is simply because we used a constant tension $T_3$ as the input to the junction conditions. What the string theory brane action tells us, is that this is not quite the whole story. The brane carries charge through dissolved flux, and this shows up in the equation of state of the brane. This stringy insight can be used to improve the junction conditions. One can show that simply replacing $T_3$ by $T_3 \sqrt{1+\frac{k^2 J_c^2}{a^6}}$ turns (\ref{eq:FinalfriedmannfromJC}) into (\ref{eq:potentialfromH}). 

The linear term in $J_c$ of (\ref{eq:potentialfromH}) is easily identified in (\ref{eq:FinalfriedmannfromJC}). We immediately see that
\begin{equation}\label{eq:JctoTheta}
	J_{c} = -\frac{\kappa_{5}\: \Delta\theta}{k \: \Delta k}.
\end{equation}

With this, the matching is complete. We have proved that the Hamiltonian's EoM and those coming from the junction conditions across the brane are unambiguously the same expression described using different variables that can straightforwardly be related. This implies that the top-down construction from the DBI-WZ action (\ref{eq:D3action}) in $10D$ from string theory yields exactly the same result as the junction condition for the co-dimension 1 shell in the five dimensional $AdS_{5}$ bulk. Hence, we have just found a string theory realisation of the dark bubble model.

\section{Four dimensional cosmology on a D3 brane}\label{sec:Cosmology}

As we have already observed, the tension of the brane is critical (to leading order), and as a consequence the cosmological constant is vanishing. Let us see why.

The junction conditions are formulated within 5D gravity, and one should work in units so that $G_5=\frac{\pi L^3}{2 N_c^2}$ is  a constant that does no change across the shell. The leading piece of the junction conditions at large scale factor $a$ then becomes
\begin{equation}\label{eq: InitialTension}
	T_3=\frac{3\Delta k}{8 \pi G_5} =-\frac{3}{8 \pi G_5} \frac{\Delta L}{L^2}.
\end{equation}
From $\Delta G_5=0$ it follows that:
\begin{equation}\label{eq:RelationLtoN}
	\frac{\Delta L}{L} = \frac{2}{3} \frac{\Delta N_c}{N_c} ,
\end{equation}
and together with (\ref{eq: InitialTension}) and $L^4=4\pi g_s N_c l_s^4$ one then obtains:
\begin{equation}\label{eq:BraneTension}
	T_3=-\frac{1}{(2\pi)^3 g_s l_s^4} \Delta N_c = -\Delta N_c T_{D3},
\end{equation}
where $\Delta N_c = -1$ corresponds to the nucleation of a single brane. Hence, we confirm that the critical tension is exactly that of a D3-brane, and that the cosmological constant will be zero. Let us now study the dynamics of the induced four dimensional cosmology.

We begin by simplifying the Friedmann equation (\ref{eq:potentialfromH}) by rewriting the mass $\mu$ and charge $\theta$ in terms of the horizons. Using (\ref{eq:muthetatohorizons}) we find:
\begin{equation}\label{eq:PotentialExtremalSubs}
	\begin{split}
		\dot{a}^{2} = -\frac{k^{2} (a_H^2 - a^2) (a_h^2 - a^2) \left(a_h^2 + a_H^2 + \tfrac{1}{k^2} + a^2\right)}{a^4} +\frac{k^{2}\:\left(a^4-a_{H}^{4}+ \frac{a_{h}\: J_{c}\sqrt{1+ \left(k^2 a_{H}^2 + a_{h}^{2}\right)}}{a_{H}} \left(1-\frac{a_{H}^2}{a^2}\right)\right)^2}{k^{2} J_{c}^2+a^6},
	\end{split}
\end{equation}
where $a_{h}$, $a_{H}$ represent the inner and outer horizons, respectively. Note that $a_{h} = a_{H}$ implies a degenerated double horizon and one can recover the extremal description.

\begin{figure}[h!]
	\centering
	\includegraphics[width=12cm]{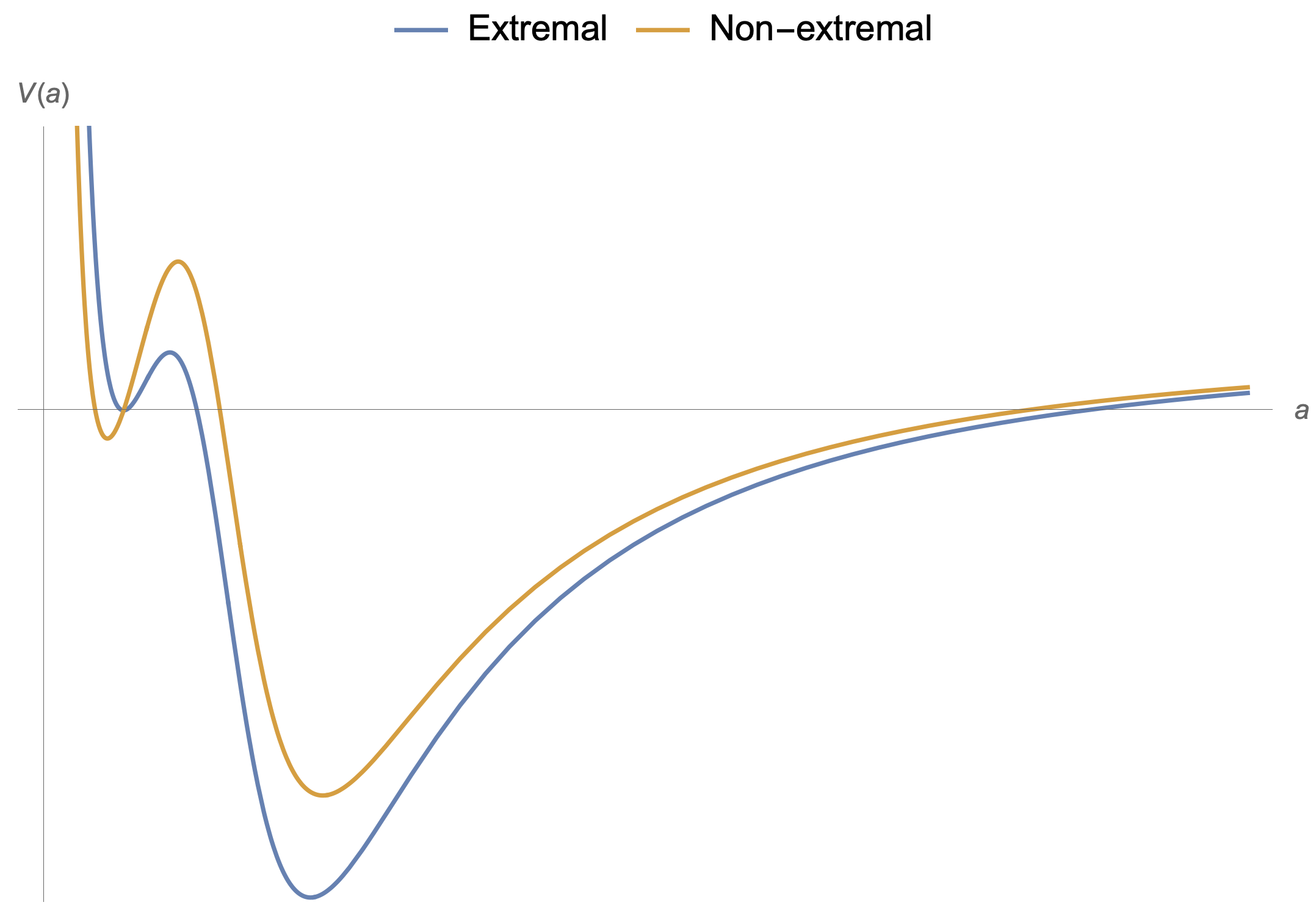}
	\caption{The potential controlling the dynamics of the embedded four dimensional universe.}
	\label{fig:Fullpotential}
\end{figure}

Equipped with this handy expression, we can investigate the potential $V$ (i.e $-\dot{a}^{2}$) controlling the dynamics of the bubble. The qualitative behaviour of $V$ can be found in figure (\ref{fig:Fullpotential}).
There are three (four) special points in the plot:
\begin{itemize}
	\item The horizons $\{a_{h}, a_{H}\}$ (first two roots in the non-extremal case), which correspond to the inner and outer horizons of the Reisser-Nordström black hole. In the extremal case, both horizon coincide, which corresponds to the first root of the blue plot. 
	\item The nucleation point $\{a_{nuc}\}$ ($2^{nd}$ or $3^{rd}$ roots, respectively), which is the value of $a$ at which the $D3$ brane (hence the $4D$ universe) will nucleate. The region between $a_{H}$ and $a_{nuc}$ is the classical forbidden region that the brane will have to tunnel through.
	\item The maximal size $\{a_{b}\}$ is given by the last root of the potential $V$. At this point, the size of the bubble is maximal, and the $4D$ universe with no $\Lambda_{4D}$ and positive curvature, will reach its maximum size and start contracting.
\end{itemize}

The potential that we have found lacks a sustained cosmological constant and experiences only a fraction of an e-folding just after creation. The expansion rate vanishes at nucleation, the universe then accelerates for a short time after which it enters a decelerated phase. Since the curvature is positive, and there is no asymptotic cosmological constant, the universe will eventually stop expanding and start to contract, entering an oscillating phase with repeated bounces.

Looking at the potential in figure \ref{fig:Fullpotential}, it is easy to see how the different parameters control the shape of the potential. By construction, the potential (\ref{eq:PotentialExtremalSubs}) goes to 1 when the scale factor $a\rightarrow \infty$. The AdS scale $k^{-1}$ will control the overall value of the potential, while $J_{c}$, which is the angular momentum of the brane with respect to the $10D$ background, will regulate the barrier of the potential. 

\begin{figure}[!htb]
	\begin{minipage}{0.48\textwidth}
		\centering
		\includegraphics[width=.8\linewidth]{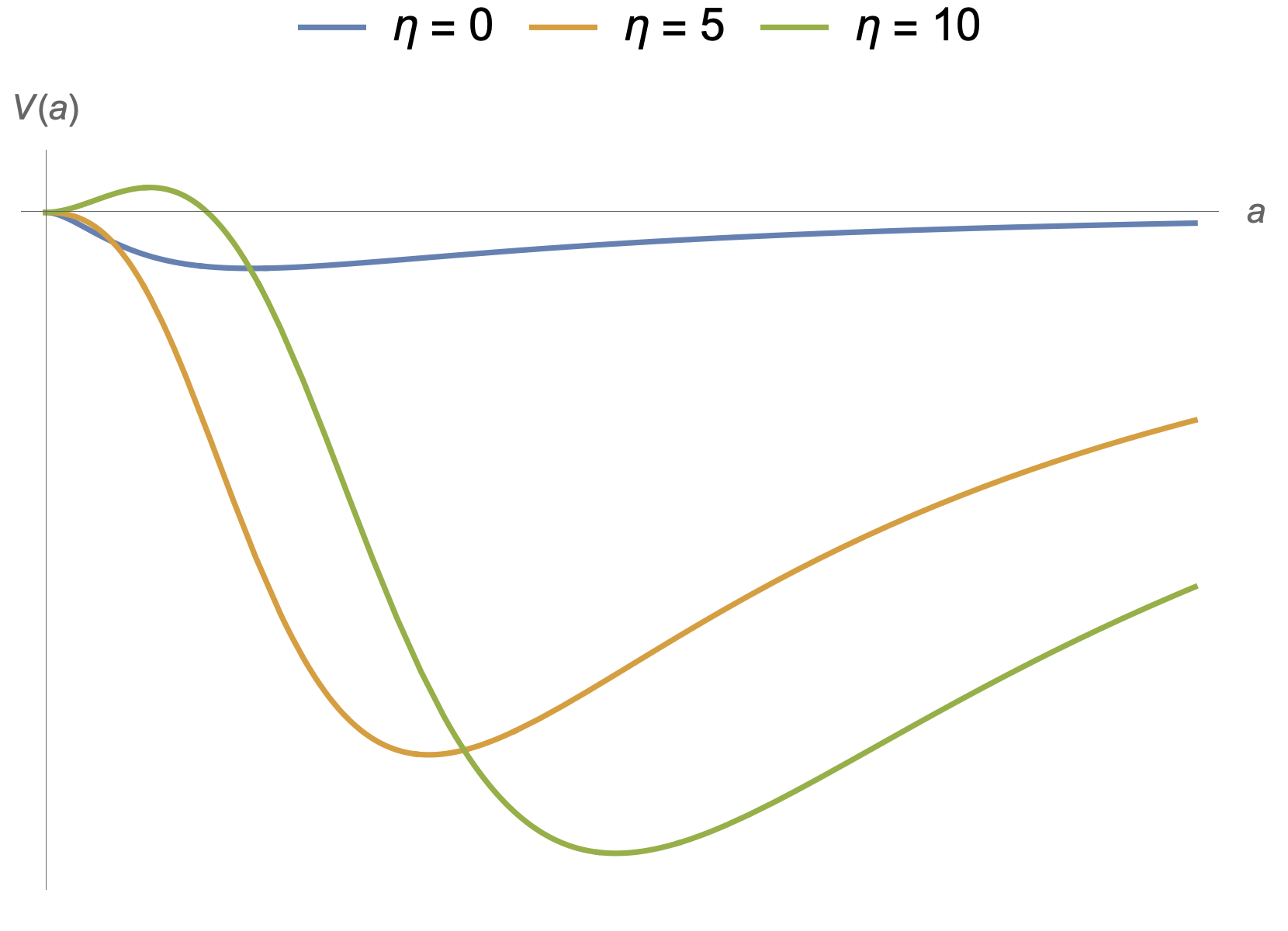}
		\caption{$T=0$ case (extremal).}\label{Fig:T0case}
	\end{minipage}\hfill
	\begin{minipage}{0.48\textwidth}
		\centering
		\includegraphics[width=.8\linewidth]{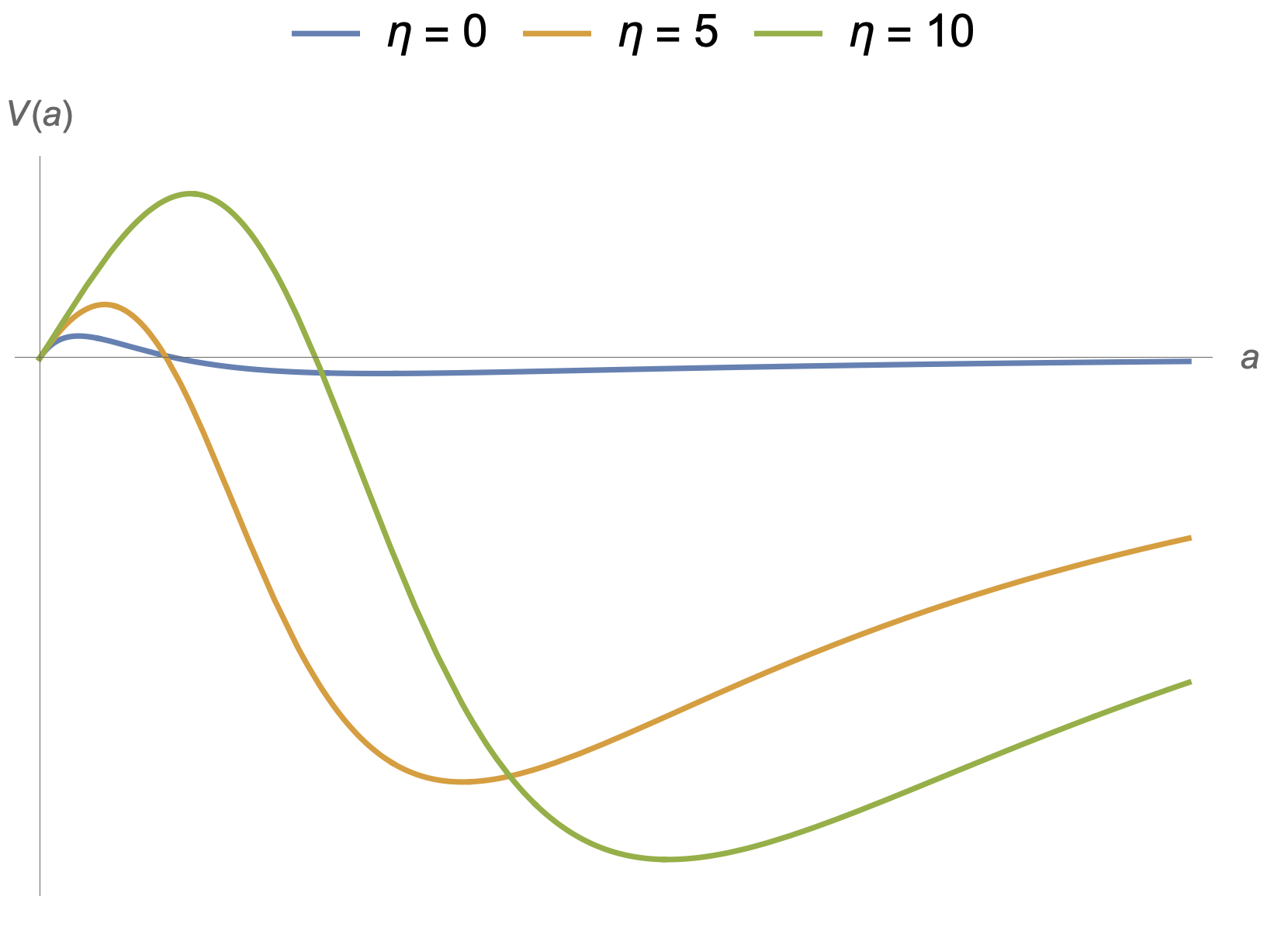}
		\caption{$T \sim 0$ case (non-degenerate).}\label{Fig:Tnon0case}
	\end{minipage}
\end{figure}

For convenience, the angular momentum $J_{c}$ can be written as $\tfrac{a_{H}^{3}}{k} \eta$, as was done in \cite{Henriksson:2019zph}, with $\eta$ as a dimensionless parameter. In figures (\ref{Fig:T0case}, \ref{Fig:Tnon0case}) we have plotted extremal as well as non-extremal potentials for several values of $\eta$.
As one can see, all values of $\eta$ in the non-degenerate black hole will yield a barrier that the brane will have to tunnel through in order to nucleate. This is not the case for the extremal black hole, as can be understood by expanding the potential close to the outer horizon $a_{H}$. Qualitatively, the expression becomes:
\begin{equation}
	V\left(a \sim a_{H}\right) \sim \left(a_{H}^{2}-a_{h}^{2}\right)\: \left(a- a_{H}\right) + f\left[a_{h}, a_{H}, k, \eta\right] \: \left(a- a_{H}\right)^{2} + \mathcal{O}\left(a - a_{H}\right)^{3}.
\end{equation}
The first term  will contribute positively in the non-extremal case, since by definition $a_{h}< a_{H}$, creating a barrier for all values of $\eta$ in the non-extremal case. On the other hand, if one imposes extremality, $a_h=a_H$, one can solve the function $f$ to find the values for $\eta$ such that there exist a barrier. 

Let us now use the relation between $J_{c}$ and $\Delta \theta$ obtained in equation (\ref{eq:JctoTheta}), together with the expression (\ref{eq:RelationLtoN}) relating the jump in the $AdS$ scale and the ratio of emitted branes $\Delta N_{c}$ to the number of branes $N_{c}$ sitting in the stack. For a single emitted brane one can then write:
\begin{equation}\label{eq:JcNtheta}
	\frac{2}{3} \frac{J_{c}}{N_{c}} k^{2} = -\kappa_{5} \: \Delta \theta.
\end{equation}
At this point, one can wonder what is the portion of charge $\Delta\theta$ that the emitted brane carries with respect to the total charge $\theta$ of the stack of branes. The most natural Ansatz is to assume that the brane carries a fraction $1/\tilde{N}$ of the total charge $\theta$, where $\tilde{N}$ is of the same order of magnitude as the total number $N_{c}$ of branes in the stack \cite{Henriksson:2019zph}. That is, the emitted brane carries it share of the total momentum or charge. So we get $\Delta \theta = -\theta/\tilde{N}$. Making use of expression (\ref{eq:muthetatohorizons}) in the large $k$ limit and extremal case, one can rewrite $\Delta \theta$ as:
\begin{equation}
	\Delta \theta = -\frac{\sqrt{2}}{\kappa_{5}\:\tilde{N}}\: k \:a_{H}^{3},
\end{equation}
Which corresponds to the RHS of eq (\ref{eq:JcNtheta}). Finally, rewriting $J_{c}$ in terms of $\eta$, one obtains:
\begin{equation}
	\eta = \frac{3}{\sqrt{2}} \frac{N_{c}}{\tilde{N}}.
\end{equation}
As long as we are off extremality, the precise value of $\eta$ (as long as it is not too different from the preferred value) does not matter too much with respect to the overall shape of the potential. At extremality, it does matter but we will not discuss that case in any further detail in this paper.

\section{Physical implications}\label{sec: Physical Implications}

Let us now discuss possible physical consequences of our embedding of the dark bubble into string theory. While there is no reason to believe that this simple example would be fully realistic, there might be some generic features that most models have in common. To find out, let us express the various physical scales in terms of the 4D Planck scale $l_4$, and the number of branes $N_c$ in the background. Ignoring numerical factors we have
\begin{equation}
	l_5^3\sim \frac{L^3}{N_c^2}, \qquad \qquad \qquad \qquad \qquad l_4^2 \sim \frac{k^2}{\Delta k} l_5^3 \sim \frac{N_c}{L}l_5^3,
\end{equation}
and find from this
\begin{equation}
	L \sim N_c^{2/3} l_5 \sim N_c^{1/2} l_4 , \qquad \qquad \qquad \qquad \qquad l_4 \sim N_c^{1/6} l_5 \ .
\end{equation}
Furthermore, from the holographic dictionary we have
\begin{equation}
	\frac{L^4}{l_s^4}\sim g_s N_c,
\end{equation}
so that
\begin{equation}
	l_{10} \sim g_s^{1/4} l_s \sim N_c^{1/4} l_4,
\end{equation}
where $l_{10}$ is the 10D Planck scale.
Hence we conclude that $L >>l_{10} >>l_4 >>l_5$ with $N_c>>1$. Having $l_5<<l_4$ is unusual and contrary to what you find in conventional compactifications, where $L>>l_5$ automatically leads to $l_5>>l_4$. The reason is the presence of the large factor $k/\Delta k  \sim N_c$ in the relation between the 5D and 4D Planck scale. As can be seen from the junction conditions, this is also the reason why a brane with a tension set by string scale can separate two AdS-vacua with curvature radius $L$ much larger than string scale (i.e. small $k$). It is a direct consequence of $l_5$ being so small.\footnote{This is very different for Randall-Sundrum constructions, \cite{Randall:1999ee,Randall:1999vf}, which are qualitatively similar to conventional compactifications. There, we have $G_{RS4}=\frac{2 k_- k_+}{k_- + k_+} G_{RS5}$, while for the dark bubble we have $G_{DB4}=\frac{2 k_- k_+}{k_- - k_+} G_{DB5}$, with the option of taking $\Delta k = k_- - k_+$ small.}  

We also note that the mass of a 4D particle induced by the endpoint of a stretched string $\frac{L}{l_s^2}=\frac{g_s^{1/2}}{l_4}$. Any particles of lower mass must then be obtained as field theory excitations in the 5D bulk that extends in the fifth dimension. This is exactly how 4D gravitational waves were shown to appear in \cite{Danielsson:2022fhd}.

By construction, there is no cosmological constant in this model since the tension of the branes is BPS. Since the backgrounds break supersymmetry one could expect, as argued by \cite{ooguri2017nonsupersymmetric} extending the WGC-conjecture, that there will be an instability so that the emitted branes actually can escape to infinity. This should take the form of a correction to the tension of the brane, making it slightly subcritical. From the dual $\mathcal{N}=4$ SYM theory, one might expect corrections of order $1/N_c^2$, since there are fields only in the adjoint representation. After the nucleation of a D3-brane, however, the gauge group is Higgsed as $SU(N)\to SU(N-1)\times U(1)$, and there are fields in the fundamental representation of the remaining $SU(N-1)$ and corrections of order $1/N_c$ may be expected. Using this, we expect a  mismatch in the junction condition of order $T_3/N_c$, yielding
\begin{equation}
	\rho_\Lambda \sim \frac{1}{L^4},
\end{equation}
corresponding to a Hubble scale
\begin{equation}
	R_H \sim N_c l_4.
\end{equation}
Matching with the present Hubble scale, which is essentially set by the cosmological constant, we find $N_c \sim 10^{60}$. This yields $L \sim 10^{-5}m$ and $l_5 \sim 10^{-45}$m, with $l_4 \sim 10^{-35}$m and $\frac{1}{l_{10}} \sim 10\:  TeV$.\footnote{If there were only $1/N_c^2$ corrections to the tension, we would instead have  $\rho_\Lambda = 1/(N_c L^4)$ and get $R_H=N_c^{3/2} l_4$. This leads to $N_c=10^{40}$, with $L\sim 10^{-15}m$ and $1/l_{10} \sim 10^9 \: GeV$. The latter intermediate scale appears in \cite{Montero:2022prj} as well. }

One can note that the size of the length scale $L$ is identical to the one of the dark dimension proposed in \cite{Montero:2022prj}, using a more conventional higher dimensional embedding. Contrary to us, they find the 5D Planck scale to be lower than the 4D-one as a natural consequence of their dimensional reduction. Intriguingly, for the dark bubble we find the 10D Planck scale to be  lowered to TeV scales (with the string scale a bit lower than that, depending on the string coupling). It is quite remarkable how a single dimensionless number, $N_c$, uniquely fixes the relation between all the scales - including the cosmological constant.

We also need to check that the induced cosmological constant is big enough to yield an eternally expanding universe. If it is too small, curvature will still win, and the universe re-contract. For $\sigma = \Delta k - \epsilon$ with $\epsilon \sim G_5/L^4 << \Delta k $, we find an extra term in the potential V.4 of the form $a^2/R_H^2$ where
\begin{equation}
	\frac{1}{R_H^2} \sim \frac{\epsilon k^2}{\Delta k} \sim \frac{G_5 k^6}{\Delta k} \sim \frac{1}{N_c^2 l_4^2} .
\end{equation}
The leading large $a$ behaviour of V.4 then becomes
\begin{equation}
	-\dot{a}^2 \sim 1 - \frac{1}{N_c^2 l_4^2} a^2 -\frac{k^2 \left(
		a_h^4 + a_h^2 a_H^2 - a_H^4 + 
		2 a_h a_H^2 \sqrt{a_h^2 + a_H^2}\eta\right)}{a^2}
\end{equation}
To make sure that the cosmological constant wins before curvature becomes important, we need, assuming $a_h \sim a_H$
\begin{equation}
	\frac{1}{N_c^2 l_4^2} a^2 \sim \frac{a_H^4}{N_c l_4^2 a^2} >>1 .
\end{equation}
This gives $a \sim N_c^{1/4} a_H$ and the constraint $a_H>>N_c^{3/4} l_4$. Note that this is consistent with $a>>R_H$, which implies that the curvature is small. It is amusing to note that this condition implies that the horizon radius of the 5D black hole sustaining our universe need to satisfy $a_H >> 10^7$km.

We conclude that if we choose $a_H$ large enough, the effective potential will be as in the figure below. We first note a big barrier that the universe must tunnel through. It then nucleates, accelerates and then enter into a phase of decelerated expansion. It then passes over the top of another hill before entering into a phase where the dark energy dominates. 

\begin{figure}[h!]
	\centering
	\includegraphics[width=10cm]{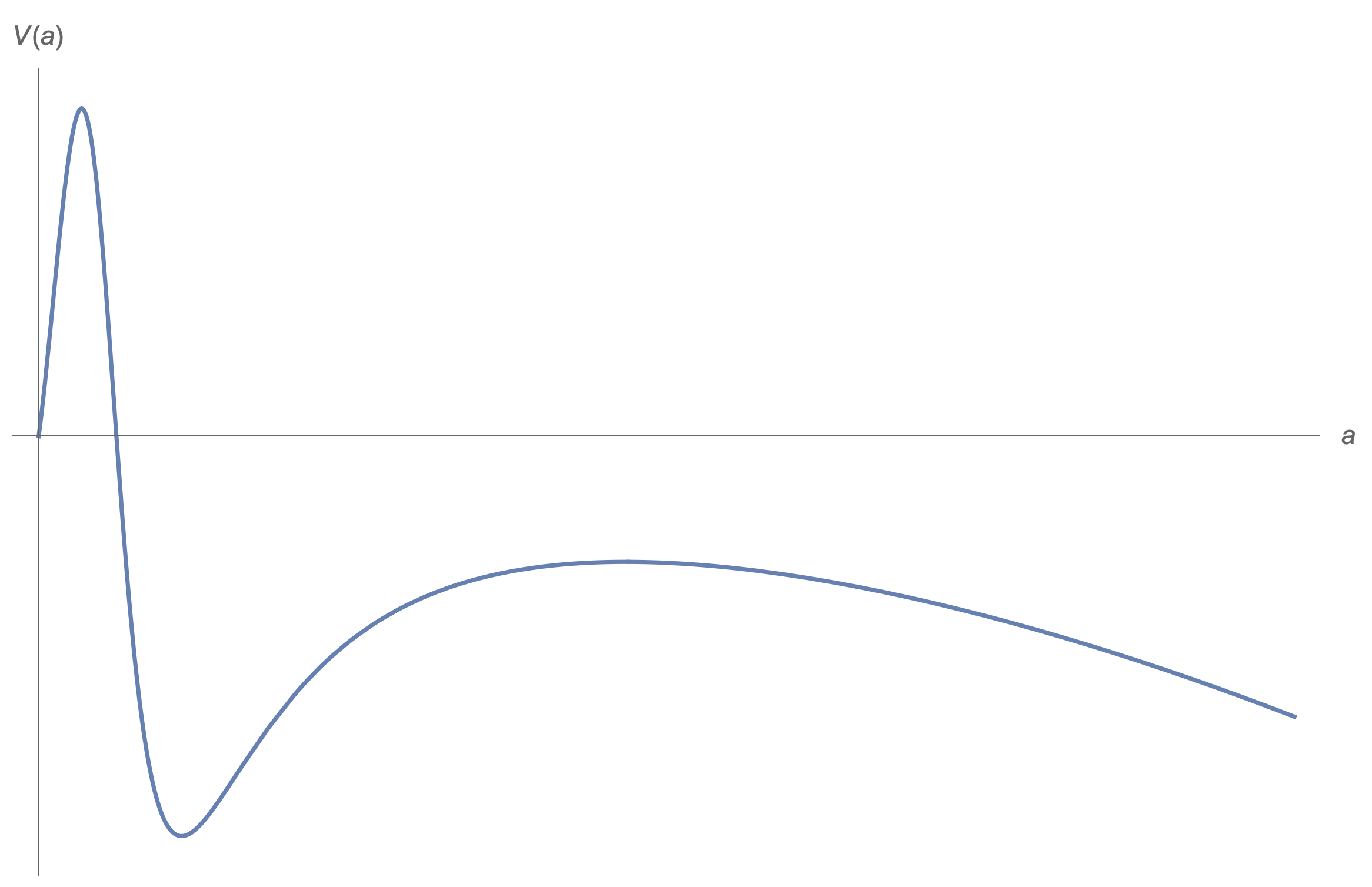}
	\caption{Using $1/N_c$ corrections, one can induce a small $\Lambda_{4D}>0$. }
	\label{fig:PotentialWcorrections}
\end{figure}

The presence of macroscopic extra dimensions is heavily constrained in conventional compactifications. The most severe astrophysical constraint on the size of the extra dimensions comes from neutron stars. One expects KK-gravitons to be emitted into the extra dimensions by supernovae, and then to be retained by the neutron stars. These clouds of KK-gravitons will emit gamma rays that in principle could be detectable. In case of six extra dimensions, \cite{Hannestad:2003yd} found that the extra dimensions must have a size smaller than about $10^{-14}$m to be consistent with observational constraints.

The constraints differ dramatically in the case of the dark bubble. The reason is that gravity in 4D is much stronger than gravity in the higher dimensions. For this reason, emission of gravitational waves into the extra dimensions is heavily suppressed. We see this from writing $G_4=\frac{2k^2}{\Delta k} G_5$ as $G_5=\frac{1}{N_c}\frac{L G_4}{2}$, were we note the huge suppression coming from $1/N_c$. As already noted in footnote 5, this makes the dark bubble unique in comparison with Randall-Sundrum and conventional compactifications.

Let us see in more detail how it works. Consider a string stretching outwards from the bubble. From the 4D point of view it will correspond to a massive particle. If the string, together with its endpoint on the brane, swings in the directions parallel with the 3 space directions of the 4D observer, there will be gravitational waves in the induced metric on the brane as well as in the 5D bulk. This uplift from 4D to 5D was explained in detail in \cite{Danielsson:2022fhd}. From the 4D perspective, the waves will be caused by the motion of the point particle corresponding to the endpoint of the string with a strength set by $G_4$. In fact, as explained in \cite{Banerjee:2020wov}, it is the backreaction of the brane to this motion that will dominate the generation of gravitational waves also in 5D. Let us now imagine that the string is swinging instead in the dimensions {\it transverse} to the brane without any motion of the endpoint on the brane. Since there is now no direct involvement of the brane in sourcing the waves, the strength of the waves is set by the strength of the gravitational force in the bulk. 

To correct the observational constraints so they become applicable to the dark bubble, we consider the relation between the 10D Planck scale and the 4D one  given by
\begin{equation}
	\frac{1}{l_4^2} \sim \frac{1}{N_c} \frac{L^6}{l_{10}^8},
\end{equation}
where we see the extra factor $1/N_c$ coming from the junction conditions. Formally absorbing the factor $1/N_c \sim 10^{-60}$ above into $L$, we find that the constraint from neutron stars is changed to  $L < 10^{-4}$m. Remarkably, this is consistent with the other estimates above. 

It would be interesting to perform a more careful analysis of how the analogue of KK-gravitons can be emitted in the dark bubble model, to establish whether the observational constraints are actually satisfied. 

It remains a challenge to see whether the presence of these other large dimensions give rise to other constraints, in particular when it comes to non-gravitational physics. To analyze this, the dark bubble model must be developed further to capture particle physics. As discussed in \cite{Danielsson:2022fhd}, fluxes within the brane come together with 2-form fields in the bulk. The interplay between these fields, and the bending of the brane, should reproduce not only electromagnetism but also the other gauge groups of the standard model. To achieve this, one would expect there to be several closely separated branes, possibly corresponding to distinct nucleation events\footnote{We thank Ivano Basile for emphasizing the latter point.}.

\section{Conclusions}\label{sec: Conclusions}

In this paper we have provided a consistent embedding of the dark bubble model of the expanding universe into string theory. The background we use corresponds to an AdS-Reissner-Nordström black hole in 5D. The charge of the black hole is due to D3-branes rapidly rotating on the internal 5-sphere. The black hole can reduce its charge by emitting one of the spinning branes through nucleation. This corresponds to the creation of a dark bubble 4D cosmology as the brane emerges after tunneling and begins to expand. Supersymmetry is broken due to the presence of the spinning branes, but to lowest order the tension of the branes remains the critical one. As a consequence, the 4D cosmological constant vanishes, and the universe will eventually reach its maximal size and then start to contract again. As a result you get a bouncing cosmology.

If one takes the WGC into account, this is not the whole story. One then expects there to be a small reduction of the tension below the critical one to guarantee that the black hole can decay by emitting branes, which escape to infinity. We have conjectured that such an effect should be present through $1/N_c$-corrections. If present, these will generate a small and positive cosmological constant, which eventually will dominate the expansion. One could argue that the presence of asymptotic supersymmetry may imply a dark bubble version of the de Sitter swampland conjecture \cite{Obied:2018sgi, Andriot_2018}. This, however, seems to be in contradiction with the expectations from WGC.

We have estimated the size of the cosmological constant obtaining intriguing results, where all important scales - from the 4D cosmological constant to the five dimensional Planck scale - are related by a single (large) integer: the number of background branes $N_c$. The measured cosmological constant requires $N_c \sim 10^{60}$, and the model predicts large extra dimensions of size $L \sim 10^{-4}$m. A preliminary analysis, making use of unique features of the dark bubble, suggests that the model is compatible with astrophysical constraints. We also note a possible parallel with the dark dimension proposed in \cite{Montero:2022prj}. To determine whether this particular model is compatible with phenomenology, it will be important to better understand how particle physics is realized.

Clearly, there could be other ways to generate a positive cosmological constant. One possibility would be to start out in a model where supersymmetry is already broken before the black hole is added \cite{Basile:2021vxh}. If the tension of the branes is supercritical, one would expect them to nucleate and mediate a decay. Another intriguing possibility is the brane/jet instabilities of \cite{Bena:2020xxb}, which are diagnosed by a similar probe brane computation as the one we performed in the current paper. In both cases, the specifics of the internal manifold is responsible for the instability; in the spinning brane backgrounds it is due to rotation, while in \cite{Bena:2020xxb} it is due to a non-trivial warp factor.

\begin{acknowledgments}
We would like to thank Miguel Montero and Ivano Basile for stimulating discussions. In addition, we would like to extend our gratitude to Rob Tielemans and Thomas van Riet, who contributed with invaluable ideas and enlightening discussions in the early stages of this work. The work of OH was supported by the Academy of Finland (grant number 330346) and the Ruth and Nils-Erik Stenb\"ack foundation. DP would like to thank the Centre for Interdisciplinary Mathematics (CIM) for financial support.
\end{acknowledgments}

\bibliography{refs}

\begin{thebibliography}{34}%
\makeatletter
\providecommand \@ifxundefined [1]{%
 \@ifx{#1\undefined}
}%
\providecommand \@ifnum [1]{%
 \ifnum #1\expandafter \@firstoftwo
 \else \expandafter \@secondoftwo
 \fi
}%
\providecommand \@ifx [1]{%
 \ifx #1\expandafter \@firstoftwo
 \else \expandafter \@secondoftwo
 \fi
}%
\providecommand \natexlab [1]{#1}%
\providecommand \enquote  [1]{``#1''}%
\providecommand \bibnamefont  [1]{#1}%
\providecommand \bibfnamefont [1]{#1}%
\providecommand \citenamefont [1]{#1}%
\providecommand \href@noop [0]{\@secondoftwo}%
\providecommand \href [0]{\begingroup \@sanitize@url \@href}%
\providecommand \@href[1]{\@@startlink{#1}\@@href}%
\providecommand \@@href[1]{\endgroup#1\@@endlink}%
\providecommand \@sanitize@url [0]{\catcode `\\12\catcode `\$12\catcode
  `\&12\catcode `\#12\catcode `\^12\catcode `\_12\catcode `\%12\relax}%
\providecommand \@@startlink[1]{}%
\providecommand \@@endlink[0]{}%
\providecommand \url  [0]{\begingroup\@sanitize@url \@url }%
\providecommand \@url [1]{\endgroup\@href {#1}{\urlprefix }}%
\providecommand \urlprefix  [0]{URL }%
\providecommand \Eprint [0]{\href }%
\providecommand \doibase [0]{https://doi.org/}%
\providecommand \selectlanguage [0]{\@gobble}%
\providecommand \bibinfo  [0]{\@secondoftwo}%
\providecommand \bibfield  [0]{\@secondoftwo}%
\providecommand \translation [1]{[#1]}%
\providecommand \BibitemOpen [0]{}%
\providecommand \bibitemStop [0]{}%
\providecommand \bibitemNoStop [0]{.\EOS\space}%
\providecommand \EOS [0]{\spacefactor3000\relax}%
\providecommand \BibitemShut  [1]{\csname bibitem#1\endcsname}%
\let\auto@bib@innerbib\@empty
\bibitem [{\citenamefont {Danielsson}\ and\ \citenamefont
  {Van~Riet}(2018)}]{Danielsson:2018ztv}%
  \BibitemOpen
  \bibfield  {author} {\bibinfo {author} {\bibfnamefont {U.~H.}\ \bibnamefont
  {Danielsson}}\ and\ \bibinfo {author} {\bibfnamefont {T.}~\bibnamefont
  {Van~Riet}},\ }\bibfield  {title} {\bibinfo {title} {{What if string theory
  has no de Sitter vacua?}},\ }\href
  {https://doi.org/10.1142/S0218271818300070} {\bibfield  {journal} {\bibinfo
  {journal} {Int. J. Mod. Phys. D}\ }\textbf {\bibinfo {volume} {27}},\
  \bibinfo {pages} {1830007} (\bibinfo {year} {2018})},\ \Eprint
  {https://arxiv.org/abs/1804.01120} {arXiv:1804.01120 [hep-th]} \BibitemShut
  {NoStop}%
\bibitem [{\citenamefont {Obied}\ \emph {et~al.}(2018)\citenamefont {Obied},
  \citenamefont {Ooguri}, \citenamefont {Spodyneiko},\ and\ \citenamefont
  {Vafa}}]{Obied:2018sgi}%
  \BibitemOpen
  \bibfield  {author} {\bibinfo {author} {\bibfnamefont {G.}~\bibnamefont
  {Obied}}, \bibinfo {author} {\bibfnamefont {H.}~\bibnamefont {Ooguri}},
  \bibinfo {author} {\bibfnamefont {L.}~\bibnamefont {Spodyneiko}},\ and\
  \bibinfo {author} {\bibfnamefont {C.}~\bibnamefont {Vafa}},\ }\bibfield
  {title} {\bibinfo {title} {{De Sitter Space and the Swampland}},\ }\href@noop
  {} {\  (\bibinfo {year} {2018})},\ \Eprint {https://arxiv.org/abs/1806.08362}
  {arXiv:1806.08362 [hep-th]} \BibitemShut {NoStop}%
\bibitem [{\citenamefont {van Beest}\ \emph {et~al.}(2021)\citenamefont {van
  Beest}, \citenamefont {Calderón-Infante}, \citenamefont {Mirfendereski},\
  and\ \citenamefont {Valenzuela}}]{vanbeest2021lectures}%
  \BibitemOpen
  \bibfield  {author} {\bibinfo {author} {\bibfnamefont {M.}~\bibnamefont {van
  Beest}}, \bibinfo {author} {\bibfnamefont {J.}~\bibnamefont
  {Calderón-Infante}}, \bibinfo {author} {\bibfnamefont {D.}~\bibnamefont
  {Mirfendereski}},\ and\ \bibinfo {author} {\bibfnamefont {I.}~\bibnamefont
  {Valenzuela}},\ }\href@noop {} {\bibinfo {title} {Lectures on the swampland
  program in string compactifications}} (\bibinfo {year} {2021}),\ \Eprint
  {https://arxiv.org/abs/2102.01111} {arXiv:2102.01111 [hep-th]} \BibitemShut
  {NoStop}%
\bibitem [{\citenamefont {Banerjee}\ \emph {et~al.}(2018)\citenamefont
  {Banerjee}, \citenamefont {Danielsson}, \citenamefont {Dibitetto},
  \citenamefont {Giri},\ and\ \citenamefont {Schillo}}]{Banerjee:2018qey}%
  \BibitemOpen
  \bibfield  {author} {\bibinfo {author} {\bibfnamefont {S.}~\bibnamefont
  {Banerjee}}, \bibinfo {author} {\bibfnamefont {U.}~\bibnamefont
  {Danielsson}}, \bibinfo {author} {\bibfnamefont {G.}~\bibnamefont
  {Dibitetto}}, \bibinfo {author} {\bibfnamefont {S.}~\bibnamefont {Giri}},\
  and\ \bibinfo {author} {\bibfnamefont {M.}~\bibnamefont {Schillo}},\
  }\bibfield  {title} {\bibinfo {title} {{Emergent de Sitter Cosmology from
  Decaying Anti\textendash{}de Sitter Space}},\ }\href
  {https://doi.org/10.1103/PhysRevLett.121.261301} {\bibfield  {journal}
  {\bibinfo  {journal} {Phys. Rev. Lett.}\ }\textbf {\bibinfo {volume} {121}},\
  \bibinfo {pages} {261301} (\bibinfo {year} {2018})},\ \Eprint
  {https://arxiv.org/abs/1807.01570} {arXiv:1807.01570 [hep-th]} \BibitemShut
  {NoStop}%
\bibitem [{\citenamefont {Banerjee}\ \emph {et~al.}(2019)\citenamefont
  {Banerjee}, \citenamefont {Danielsson}, \citenamefont {Dibitetto},
  \citenamefont {Giri},\ and\ \citenamefont {Schillo}}]{Banerjee:2019fzz}%
  \BibitemOpen
  \bibfield  {author} {\bibinfo {author} {\bibfnamefont {S.}~\bibnamefont
  {Banerjee}}, \bibinfo {author} {\bibfnamefont {U.}~\bibnamefont
  {Danielsson}}, \bibinfo {author} {\bibfnamefont {G.}~\bibnamefont
  {Dibitetto}}, \bibinfo {author} {\bibfnamefont {S.}~\bibnamefont {Giri}},\
  and\ \bibinfo {author} {\bibfnamefont {M.}~\bibnamefont {Schillo}},\
  }\bibfield  {title} {\bibinfo {title} {{de Sitter Cosmology on an expanding
  bubble}},\ }\href {https://doi.org/10.1007/JHEP10(2019)164} {\bibfield
  {journal} {\bibinfo  {journal} {JHEP}\ }\textbf {\bibinfo {volume} {10}},\
  \bibinfo {pages} {164}},\ \Eprint {https://arxiv.org/abs/1907.04268}
  {arXiv:1907.04268 [hep-th]} \BibitemShut {NoStop}%
\bibitem [{\citenamefont {Banerjee}\ \emph
  {et~al.}(2020{\natexlab{a}})\citenamefont {Banerjee}, \citenamefont
  {Danielsson},\ and\ \citenamefont {Giri}}]{Banerjee:2020wix}%
  \BibitemOpen
  \bibfield  {author} {\bibinfo {author} {\bibfnamefont {S.}~\bibnamefont
  {Banerjee}}, \bibinfo {author} {\bibfnamefont {U.}~\bibnamefont
  {Danielsson}},\ and\ \bibinfo {author} {\bibfnamefont {S.}~\bibnamefont
  {Giri}},\ }\bibfield  {title} {\bibinfo {title} {{Dark bubbles: decorating
  the wall}},\ }\href {https://doi.org/10.1007/JHEP04(2020)085} {\bibfield
  {journal} {\bibinfo  {journal} {JHEP}\ }\textbf {\bibinfo {volume} {04}},\
  \bibinfo {pages} {085}},\ \Eprint {https://arxiv.org/abs/2001.07433}
  {arXiv:2001.07433 [hep-th]} \BibitemShut {NoStop}%
\bibitem [{\citenamefont {Banerjee}\ \emph
  {et~al.}(2020{\natexlab{b}})\citenamefont {Banerjee}, \citenamefont
  {Danielsson},\ and\ \citenamefont {Giri}}]{Banerjee:2020wov}%
  \BibitemOpen
  \bibfield  {author} {\bibinfo {author} {\bibfnamefont {S.}~\bibnamefont
  {Banerjee}}, \bibinfo {author} {\bibfnamefont {U.}~\bibnamefont
  {Danielsson}},\ and\ \bibinfo {author} {\bibfnamefont {S.}~\bibnamefont
  {Giri}},\ }\bibfield  {title} {\bibinfo {title} {{Bubble needs strings}},\
  }\href {https://doi.org/10.1007/JHEP03(2021)250} {\bibfield  {journal}
  {\bibinfo  {journal} {JHEP}\ }\textbf {\bibinfo {volume} {21}},\ \bibinfo
  {pages} {250}},\ \Eprint {https://arxiv.org/abs/2009.01597} {arXiv:2009.01597
  [hep-th]} \BibitemShut {NoStop}%
\bibitem [{\citenamefont {Banerjee}\ \emph
  {et~al.}(2021{\natexlab{a}})\citenamefont {Banerjee}, \citenamefont
  {Danielsson},\ and\ \citenamefont {Giri}}]{Banerjee:2021qei}%
  \BibitemOpen
  \bibfield  {author} {\bibinfo {author} {\bibfnamefont {S.}~\bibnamefont
  {Banerjee}}, \bibinfo {author} {\bibfnamefont {U.}~\bibnamefont
  {Danielsson}},\ and\ \bibinfo {author} {\bibfnamefont {S.}~\bibnamefont
  {Giri}},\ }\bibfield  {title} {\bibinfo {title} {{Dark bubbles and black
  holes}},\ }\href {https://doi.org/10.1007/JHEP09(2021)158} {\bibfield
  {journal} {\bibinfo  {journal} {JHEP}\ }\textbf {\bibinfo {volume} {09}},\
  \bibinfo {pages} {158}},\ \Eprint {https://arxiv.org/abs/2102.02164}
  {arXiv:2102.02164 [hep-th]} \BibitemShut {NoStop}%
\bibitem [{\citenamefont {Banerjee}\ \emph
  {et~al.}(2021{\natexlab{b}})\citenamefont {Banerjee}, \citenamefont
  {Danielsson},\ and\ \citenamefont {Giri}}]{Banerjee:2021yrb}%
  \BibitemOpen
  \bibfield  {author} {\bibinfo {author} {\bibfnamefont {S.}~\bibnamefont
  {Banerjee}}, \bibinfo {author} {\bibfnamefont {U.}~\bibnamefont
  {Danielsson}},\ and\ \bibinfo {author} {\bibfnamefont {S.}~\bibnamefont
  {Giri}},\ }\bibfield  {title} {\bibinfo {title} {{Curing with Hemlock:
  Escaping the swampland using instabilities from string theory}},\ }\href
  {https://doi.org/10.1142/S0218271821420293} {\bibfield  {journal} {\bibinfo
  {journal} {Int. J. Mod. Phys. D}\ }\textbf {\bibinfo {volume} {30}},\
  \bibinfo {pages} {2142029} (\bibinfo {year} {2021}{\natexlab{b}})},\ \Eprint
  {https://arxiv.org/abs/2103.17121} {arXiv:2103.17121 [hep-th]} \BibitemShut
  {NoStop}%
\bibitem [{\citenamefont {Ooguri}\ and\ \citenamefont
  {Vafa}(2017)}]{ooguri2017nonsupersymmetric}%
  \BibitemOpen
  \bibfield  {author} {\bibinfo {author} {\bibfnamefont {H.}~\bibnamefont
  {Ooguri}}\ and\ \bibinfo {author} {\bibfnamefont {C.}~\bibnamefont {Vafa}},\
  }\href@noop {} {\bibinfo {title} {Non-supersymmetric ads and the swampland}}
  (\bibinfo {year} {2017}),\ \Eprint {https://arxiv.org/abs/1610.01533}
  {arXiv:1610.01533 [hep-th]} \BibitemShut {NoStop}%
\bibitem [{\citenamefont {Freivogel}\ and\ \citenamefont
  {Kleban}(2016)}]{freivogel2016vacua}%
  \BibitemOpen
  \bibfield  {author} {\bibinfo {author} {\bibfnamefont {B.}~\bibnamefont
  {Freivogel}}\ and\ \bibinfo {author} {\bibfnamefont {M.}~\bibnamefont
  {Kleban}},\ }\href@noop {} {\bibinfo {title} {Vacua morghulis}} (\bibinfo
  {year} {2016}),\ \Eprint {https://arxiv.org/abs/1610.04564} {arXiv:1610.04564
  [hep-th]} \BibitemShut {NoStop}%
\bibitem [{\citenamefont {Danielsson}\ and\ \citenamefont
  {Dibitetto}(2017)}]{2016mtxDanielsson}%
  \BibitemOpen
  \bibfield  {author} {\bibinfo {author} {\bibfnamefont {U.}~\bibnamefont
  {Danielsson}}\ and\ \bibinfo {author} {\bibfnamefont {G.}~\bibnamefont
  {Dibitetto}},\ }\bibfield  {title} {\bibinfo {title} {{Fate of stringy AdS
  vacua and the weak gravity conjecture}},\ }\href
  {https://doi.org/10.1103/PhysRevD.96.026020} {\bibfield  {journal} {\bibinfo
  {journal} {Phys. Rev. D}\ }\textbf {\bibinfo {volume} {96}},\ \bibinfo
  {pages} {026020} (\bibinfo {year} {2017})},\ \Eprint
  {https://arxiv.org/abs/1611.01395} {arXiv:1611.01395 [hep-th]} \BibitemShut
  {NoStop}%
\bibitem [{\citenamefont {Danielsson}\ \emph {et~al.}(2021)\citenamefont
  {Danielsson}, \citenamefont {Panizo}, \citenamefont {Tielemans},\ and\
  \citenamefont {Van~Riet}}]{Danielsson:2021tyb}%
  \BibitemOpen
  \bibfield  {author} {\bibinfo {author} {\bibfnamefont {U.~H.}\ \bibnamefont
  {Danielsson}}, \bibinfo {author} {\bibfnamefont {D.}~\bibnamefont {Panizo}},
  \bibinfo {author} {\bibfnamefont {R.}~\bibnamefont {Tielemans}},\ and\
  \bibinfo {author} {\bibfnamefont {T.}~\bibnamefont {Van~Riet}},\ }\bibfield
  {title} {\bibinfo {title} {{Higher-dimensional view on quantum cosmology}},\
  }\href {https://doi.org/10.1103/PhysRevD.104.086015} {\bibfield  {journal}
  {\bibinfo  {journal} {Phys. Rev. D}\ }\textbf {\bibinfo {volume} {104}},\
  \bibinfo {pages} {086015} (\bibinfo {year} {2021})},\ \Eprint
  {https://arxiv.org/abs/2105.03253} {arXiv:2105.03253 [hep-th]} \BibitemShut
  {NoStop}%
\bibitem [{\citenamefont {Basile}\ and\ \citenamefont
  {Lanza}(2020)}]{Basile:2020mpt}%
  \BibitemOpen
  \bibfield  {author} {\bibinfo {author} {\bibfnamefont {I.}~\bibnamefont
  {Basile}}\ and\ \bibinfo {author} {\bibfnamefont {S.}~\bibnamefont {Lanza}},\
  }\bibfield  {title} {\bibinfo {title} {{de Sitter in non-supersymmetric
  string theories: no-go theorems and brane-worlds}},\ }\href
  {https://doi.org/10.1007/JHEP10(2020)108} {\bibfield  {journal} {\bibinfo
  {journal} {JHEP}\ }\textbf {\bibinfo {volume} {10}},\ \bibinfo {pages}
  {108}},\ \Eprint {https://arxiv.org/abs/2007.13757} {arXiv:2007.13757
  [hep-th]} \BibitemShut {NoStop}%
\bibitem [{\citenamefont {Berglund}\ \emph {et~al.}(2021)\citenamefont
  {Berglund}, \citenamefont {Hübsch},\ and\ \citenamefont
  {Minic}}]{Berglund2021}%
  \BibitemOpen
  \bibfield  {author} {\bibinfo {author} {\bibfnamefont {P.}~\bibnamefont
  {Berglund}}, \bibinfo {author} {\bibfnamefont {T.}~\bibnamefont {Hübsch}},\
  and\ \bibinfo {author} {\bibfnamefont {D.}~\bibnamefont {Minic}},\ }\bibfield
   {title} {\bibinfo {title} {Stringy bubbles solve de sitter troubles},\
  }\href {https://doi.org/10.3390/universe7100363} {\bibfield  {journal}
  {\bibinfo  {journal} {Universe}\ }\textbf {\bibinfo {volume} {7}},\ \bibinfo
  {pages} {363} (\bibinfo {year} {2021})},\ \Eprint
  {https://arxiv.org/abs/2109.01122} {arXiv:2109.01122 [hep-th]} \BibitemShut
  {NoStop}%
\bibitem [{\citenamefont {Cvetic}\ and\ \citenamefont
  {Gubser}(1999{\natexlab{a}})}]{Cvetic_1999}%
  \BibitemOpen
  \bibfield  {author} {\bibinfo {author} {\bibfnamefont {M.}~\bibnamefont
  {Cvetic}}\ and\ \bibinfo {author} {\bibfnamefont {S.~S.}\ \bibnamefont
  {Gubser}},\ }\bibfield  {title} {\bibinfo {title} {Phases of r-charged black
  holes, spinning branes and strongly coupled gauge theories},\ }\href
  {https://doi.org/10.1088/1126-6708/1999/04/024} {\bibfield  {journal}
  {\bibinfo  {journal} {Journal of High Energy Physics}\ }\textbf {\bibinfo
  {volume} {1999}},\ \bibinfo {pages} {024} (\bibinfo {year}
  {1999}{\natexlab{a}})}\BibitemShut {NoStop}%
\bibitem [{\citenamefont {Koga}\ \emph {et~al.}(2022)\citenamefont {Koga},
  \citenamefont {Oshita},\ and\ \citenamefont {Ueda}}]{Koga:2022opd}%
  \BibitemOpen
  \bibfield  {author} {\bibinfo {author} {\bibfnamefont {I.}~\bibnamefont
  {Koga}}, \bibinfo {author} {\bibfnamefont {N.}~\bibnamefont {Oshita}},\ and\
  \bibinfo {author} {\bibfnamefont {K.}~\bibnamefont {Ueda}},\ }\bibfield
  {title} {\bibinfo {title} {{dS${}_4$ universe emergent from Kerr-AdS${}_5$
  spacetime: bubble nucleation catalyzed by a black hole}},\ }\href@noop {} {\
  (\bibinfo {year} {2022})},\ \Eprint {https://arxiv.org/abs/2209.05625}
  {arXiv:2209.05625 [hep-th]} \BibitemShut {NoStop}%
\bibitem [{\citenamefont {Henriksson}\ \emph {et~al.}(2019)\citenamefont
  {Henriksson}, \citenamefont {Hoyos},\ and\ \citenamefont
  {Jokela}}]{Henriksson:2019zph}%
  \BibitemOpen
  \bibfield  {author} {\bibinfo {author} {\bibfnamefont {O.}~\bibnamefont
  {Henriksson}}, \bibinfo {author} {\bibfnamefont {C.}~\bibnamefont {Hoyos}},\
  and\ \bibinfo {author} {\bibfnamefont {N.}~\bibnamefont {Jokela}},\
  }\bibfield  {title} {\bibinfo {title} {{Novel color superconducting phases of
  $\cal{N}$ = 4 super Yang-Mills at strong coupling}},\ }\href
  {https://doi.org/10.1007/JHEP09(2019)088} {\bibfield  {journal} {\bibinfo
  {journal} {JHEP}\ }\textbf {\bibinfo {volume} {09}},\ \bibinfo {pages}
  {088}},\ \Eprint {https://arxiv.org/abs/1907.01562} {arXiv:1907.01562
  [hep-th]} \BibitemShut {NoStop}%
\bibitem [{\citenamefont {Israel}(1966)}]{Israel:1966rt}%
  \BibitemOpen
  \bibfield  {author} {\bibinfo {author} {\bibfnamefont {W.}~\bibnamefont
  {Israel}},\ }\bibfield  {title} {\bibinfo {title} {{Singular hypersurfaces
  and thin shells in general relativity}},\ }\href
  {https://doi.org/10.1007/BF02710419} {\bibfield  {journal} {\bibinfo
  {journal} {Nuovo Cim. B}\ }\textbf {\bibinfo {volume} {44S10}},\ \bibinfo
  {pages} {1} (\bibinfo {year} {1966})},\ \bibinfo {note} {[Erratum: Nuovo
  Cim.B 48, 463 (1967)]}\BibitemShut {NoStop}%
\bibitem [{\citenamefont {Danielsson}\ \emph {et~al.}(2022)\citenamefont
  {Danielsson}, \citenamefont {Panizo},\ and\ \citenamefont
  {Tielemans}}]{Danielsson:2022fhd}%
  \BibitemOpen
  \bibfield  {author} {\bibinfo {author} {\bibfnamefont {U.}~\bibnamefont
  {Danielsson}}, \bibinfo {author} {\bibfnamefont {D.}~\bibnamefont {Panizo}},\
  and\ \bibinfo {author} {\bibfnamefont {R.}~\bibnamefont {Tielemans}},\
  }\bibfield  {title} {\bibinfo {title} {{Gravitational waves in dark bubble
  cosmology}},\ }\href {https://doi.org/10.1103/PhysRevD.106.024002} {\bibfield
   {journal} {\bibinfo  {journal} {Phys. Rev. D}\ }\textbf {\bibinfo {volume}
  {106}},\ \bibinfo {pages} {024002} (\bibinfo {year} {2022})},\ \Eprint
  {https://arxiv.org/abs/2202.00545} {arXiv:2202.00545 [hep-th]} \BibitemShut
  {NoStop}%
\bibitem [{\citenamefont {Montero}\ \emph {et~al.}(2022)\citenamefont
  {Montero}, \citenamefont {Vafa},\ and\ \citenamefont
  {Valenzuela}}]{Montero:2022prj}%
  \BibitemOpen
  \bibfield  {author} {\bibinfo {author} {\bibfnamefont {M.}~\bibnamefont
  {Montero}}, \bibinfo {author} {\bibfnamefont {C.}~\bibnamefont {Vafa}},\ and\
  \bibinfo {author} {\bibfnamefont {I.}~\bibnamefont {Valenzuela}},\ }\bibfield
   {title} {\bibinfo {title} {{The Dark Dimension and the Swampland}},\
  }\href@noop {} {\  (\bibinfo {year} {2022})},\ \Eprint
  {https://arxiv.org/abs/2205.12293} {arXiv:2205.12293 [hep-th]} \BibitemShut
  {NoStop}%
\bibitem [{\citenamefont {Behrndt}\ \emph {et~al.}(1999)\citenamefont
  {Behrndt}, \citenamefont {Cveti{\v{c}}},\ and\ \citenamefont
  {Sabra}}]{Behrndt_1999}%
  \BibitemOpen
  \bibfield  {author} {\bibinfo {author} {\bibfnamefont {K.}~\bibnamefont
  {Behrndt}}, \bibinfo {author} {\bibfnamefont {M.}~\bibnamefont
  {Cveti{\v{c}}}},\ and\ \bibinfo {author} {\bibfnamefont {W.}~\bibnamefont
  {Sabra}},\ }\bibfield  {title} {\bibinfo {title} {Non-extreme black holes of
  five-dimensional n = 2 {AdS} supergravity},\ }\href
  {https://doi.org/10.1016/s0550-3213(99)00243-6} {\bibfield  {journal}
  {\bibinfo  {journal} {Nuclear Physics B}\ }\textbf {\bibinfo {volume}
  {553}},\ \bibinfo {pages} {317} (\bibinfo {year} {1999})}\BibitemShut
  {NoStop}%
\bibitem [{\citenamefont {Cvetic}\ \emph {et~al.}(1999)\citenamefont {Cvetic},
  \citenamefont {Duff}, \citenamefont {Hoxha}, \citenamefont {Liu},
  \citenamefont {Lü}, \citenamefont {Lu}, \citenamefont {Martinez-Acosta},
  \citenamefont {Pope}, \citenamefont {Sati},\ and\ \citenamefont
  {Tran}}]{Cvetic__1999}%
  \BibitemOpen
  \bibfield  {author} {\bibinfo {author} {\bibfnamefont {M.}~\bibnamefont
  {Cvetic}}, \bibinfo {author} {\bibfnamefont {M.}~\bibnamefont {Duff}},
  \bibinfo {author} {\bibfnamefont {P.}~\bibnamefont {Hoxha}}, \bibinfo
  {author} {\bibfnamefont {J.~T.}\ \bibnamefont {Liu}}, \bibinfo {author}
  {\bibfnamefont {H.}~\bibnamefont {Lü}}, \bibinfo {author} {\bibfnamefont
  {J.}~\bibnamefont {Lu}}, \bibinfo {author} {\bibfnamefont {R.}~\bibnamefont
  {Martinez-Acosta}}, \bibinfo {author} {\bibfnamefont {C.}~\bibnamefont
  {Pope}}, \bibinfo {author} {\bibfnamefont {H.}~\bibnamefont {Sati}},\ and\
  \bibinfo {author} {\bibfnamefont {T.}~\bibnamefont {Tran}},\ }\bibfield
  {title} {\bibinfo {title} {Embedding {AdS} black holes in ten and eleven
  dimensions},\ }\href {https://doi.org/10.1016/s0550-3213(99)00419-8}
  {\bibfield  {journal} {\bibinfo  {journal} {Nuclear Physics B}\ }\textbf
  {\bibinfo {volume} {558}},\ \bibinfo {pages} {96} (\bibinfo {year}
  {1999})}\BibitemShut {NoStop}%
\bibitem [{\citenamefont {Cai}\ and\ \citenamefont {Soh}(1999)}]{Cai:1998ji}%
  \BibitemOpen
  \bibfield  {author} {\bibinfo {author} {\bibfnamefont {R.-G.}\ \bibnamefont
  {Cai}}\ and\ \bibinfo {author} {\bibfnamefont {K.-S.}\ \bibnamefont {Soh}},\
  }\bibfield  {title} {\bibinfo {title} {{Critical behavior in the rotating
  D-branes}},\ }\href {https://doi.org/10.1142/S0217732399001966} {\bibfield
  {journal} {\bibinfo  {journal} {Mod. Phys. Lett. A}\ }\textbf {\bibinfo
  {volume} {14}},\ \bibinfo {pages} {1895} (\bibinfo {year} {1999})},\ \Eprint
  {https://arxiv.org/abs/hep-th/9812121} {arXiv:hep-th/9812121} \BibitemShut
  {NoStop}%
\bibitem [{\citenamefont {Cvetic}\ and\ \citenamefont
  {Gubser}(1999{\natexlab{b}})}]{Cvetic_3}%
  \BibitemOpen
  \bibfield  {author} {\bibinfo {author} {\bibfnamefont {M.}~\bibnamefont
  {Cvetic}}\ and\ \bibinfo {author} {\bibfnamefont {S.~S.}\ \bibnamefont
  {Gubser}},\ }\bibfield  {title} {\bibinfo {title} {Thermodynamic stability
  and phases of general spinning branes},\ }\href
  {https://doi.org/10.1088/1126-6708/1999/07/010} {\bibfield  {journal}
  {\bibinfo  {journal} {Journal of High Energy Physics}\ }\textbf {\bibinfo
  {volume} {1999}},\ \bibinfo {pages} {010} (\bibinfo {year}
  {1999}{\natexlab{b}})}\BibitemShut {NoStop}%
\bibitem [{\citenamefont {Chamblin}\ \emph {et~al.}(1999)\citenamefont
  {Chamblin}, \citenamefont {Emparan}, \citenamefont {Johnson},\ and\
  \citenamefont {Myers}}]{Chamblin_1999}%
  \BibitemOpen
  \bibfield  {author} {\bibinfo {author} {\bibfnamefont {A.}~\bibnamefont
  {Chamblin}}, \bibinfo {author} {\bibfnamefont {R.}~\bibnamefont {Emparan}},
  \bibinfo {author} {\bibfnamefont {C.~V.}\ \bibnamefont {Johnson}},\ and\
  \bibinfo {author} {\bibfnamefont {R.~C.}\ \bibnamefont {Myers}},\ }\bibfield
  {title} {\bibinfo {title} {Charged {AdS} black holes and catastrophic
  holography},\ }\bibfield  {journal} {\bibinfo  {journal} {Physical Review D}\
  }\textbf {\bibinfo {volume} {60}},\ \href
  {https://doi.org/10.1103/physrevd.60.064018} {10.1103/physrevd.60.064018}
  (\bibinfo {year} {1999})\BibitemShut {NoStop}%
\bibitem [{\citenamefont {Yamada}(2008)}]{Yamada:2008em}%
  \BibitemOpen
  \bibfield  {author} {\bibinfo {author} {\bibfnamefont {D.}~\bibnamefont
  {Yamada}},\ }\bibfield  {title} {\bibinfo {title} {{Fragmentation of Spinning
  Branes}},\ }\href {https://doi.org/10.1088/0264-9381/25/14/145006} {\bibfield
   {journal} {\bibinfo  {journal} {Class. Quant. Grav.}\ }\textbf {\bibinfo
  {volume} {25}},\ \bibinfo {pages} {145006} (\bibinfo {year} {2008})},\
  \Eprint {https://arxiv.org/abs/0802.3508} {arXiv:0802.3508 [hep-th]}
  \BibitemShut {NoStop}%
\bibitem [{\citenamefont {Henriksson}(2022)}]{Henriksson:2021zei}%
  \BibitemOpen
  \bibfield  {author} {\bibinfo {author} {\bibfnamefont {O.}~\bibnamefont
  {Henriksson}},\ }\bibfield  {title} {\bibinfo {title} {{Black brane
  evaporation through D-brane bubble nucleation}},\ }\href
  {https://doi.org/10.1103/PhysRevD.105.L041901} {\bibfield  {journal}
  {\bibinfo  {journal} {Phys. Rev. D}\ }\textbf {\bibinfo {volume} {105}},\
  \bibinfo {pages} {L041901} (\bibinfo {year} {2022})},\ \Eprint
  {https://arxiv.org/abs/2106.13254} {arXiv:2106.13254 [hep-th]} \BibitemShut
  {NoStop}%
\bibitem [{\citenamefont {Randall}\ and\ \citenamefont
  {Sundrum}(1999{\natexlab{a}})}]{Randall:1999ee}%
  \BibitemOpen
  \bibfield  {author} {\bibinfo {author} {\bibfnamefont {L.}~\bibnamefont
  {Randall}}\ and\ \bibinfo {author} {\bibfnamefont {R.}~\bibnamefont
  {Sundrum}},\ }\bibfield  {title} {\bibinfo {title} {{A Large mass hierarchy
  from a small extra dimension}},\ }\href
  {https://doi.org/10.1103/PhysRevLett.83.3370} {\bibfield  {journal} {\bibinfo
   {journal} {Phys. Rev. Lett.}\ }\textbf {\bibinfo {volume} {83}},\ \bibinfo
  {pages} {3370} (\bibinfo {year} {1999}{\natexlab{a}})},\ \Eprint
  {https://arxiv.org/abs/hep-ph/9905221} {arXiv:hep-ph/9905221} \BibitemShut
  {NoStop}%
\bibitem [{\citenamefont {Randall}\ and\ \citenamefont
  {Sundrum}(1999{\natexlab{b}})}]{Randall:1999vf}%
  \BibitemOpen
  \bibfield  {author} {\bibinfo {author} {\bibfnamefont {L.}~\bibnamefont
  {Randall}}\ and\ \bibinfo {author} {\bibfnamefont {R.}~\bibnamefont
  {Sundrum}},\ }\bibfield  {title} {\bibinfo {title} {{An Alternative to
  compactification}},\ }\href {https://doi.org/10.1103/PhysRevLett.83.4690}
  {\bibfield  {journal} {\bibinfo  {journal} {Phys. Rev. Lett.}\ }\textbf
  {\bibinfo {volume} {83}},\ \bibinfo {pages} {4690} (\bibinfo {year}
  {1999}{\natexlab{b}})},\ \Eprint {https://arxiv.org/abs/hep-th/9906064}
  {arXiv:hep-th/9906064} \BibitemShut {NoStop}%
\bibitem [{\citenamefont {Hannestad}\ and\ \citenamefont
  {Raffelt}(2003)}]{Hannestad:2003yd}%
  \BibitemOpen
  \bibfield  {author} {\bibinfo {author} {\bibfnamefont {S.}~\bibnamefont
  {Hannestad}}\ and\ \bibinfo {author} {\bibfnamefont {G.~G.}\ \bibnamefont
  {Raffelt}},\ }\bibfield  {title} {\bibinfo {title} {{Supernova and neutron
  star limits on large extra dimensions reexamined}},\ }\href
  {https://doi.org/10.1103/PhysRevD.69.029901} {\bibfield  {journal} {\bibinfo
  {journal} {Phys. Rev. D}\ }\textbf {\bibinfo {volume} {67}},\ \bibinfo
  {pages} {125008} (\bibinfo {year} {2003})},\ \bibinfo {note} {[Erratum:
  Phys.Rev.D 69, 029901 (2004)]},\ \Eprint
  {https://arxiv.org/abs/hep-ph/0304029} {arXiv:hep-ph/0304029} \BibitemShut
  {NoStop}%
\bibitem [{\citenamefont {Andriot}(2018)}]{Andriot_2018}%
  \BibitemOpen
  \bibfield  {author} {\bibinfo {author} {\bibfnamefont {D.}~\bibnamefont
  {Andriot}},\ }\bibfield  {title} {\bibinfo {title} {On the de sitter
  swampland criterion},\ }\href
  {https://doi.org/10.1016/j.physletb.2018.09.022} {\bibfield  {journal}
  {\bibinfo  {journal} {Physics Letters B}\ }\textbf {\bibinfo {volume}
  {785}},\ \bibinfo {pages} {570} (\bibinfo {year} {2018})}\BibitemShut
  {NoStop}%
\bibitem [{\citenamefont {Basile}(2021)}]{Basile:2021vxh}%
  \BibitemOpen
  \bibfield  {author} {\bibinfo {author} {\bibfnamefont {I.}~\bibnamefont
  {Basile}},\ }\bibfield  {title} {\bibinfo {title} {{Supersymmetry breaking
  and stability in string vacua: Brane dynamics, bubbles and the swampland}},\
  }\href {https://doi.org/10.1007/s40766-021-00024-9} {\bibfield  {journal}
  {\bibinfo  {journal} {Riv. Nuovo Cim.}\ }\textbf {\bibinfo {volume} {44}},\
  \bibinfo {pages} {499} (\bibinfo {year} {2021})},\ \Eprint
  {https://arxiv.org/abs/2107.02814} {arXiv:2107.02814 [hep-th]} \BibitemShut
  {NoStop}%
\bibitem [{\citenamefont {Bena}\ \emph {et~al.}(2020)\citenamefont {Bena},
  \citenamefont {Pilch},\ and\ \citenamefont {Warner}}]{Bena:2020xxb}%
  \BibitemOpen
  \bibfield  {author} {\bibinfo {author} {\bibfnamefont {I.}~\bibnamefont
  {Bena}}, \bibinfo {author} {\bibfnamefont {K.}~\bibnamefont {Pilch}},\ and\
  \bibinfo {author} {\bibfnamefont {N.~P.}\ \bibnamefont {Warner}},\ }\bibfield
   {title} {\bibinfo {title} {{Brane-Jet Instabilities}},\ }\href
  {https://doi.org/10.1007/JHEP10(2020)091} {\bibfield  {journal} {\bibinfo
  {journal} {JHEP}\ }\textbf {\bibinfo {volume} {10}},\ \bibinfo {pages}
  {091}},\ \Eprint {https://arxiv.org/abs/2003.02851} {arXiv:2003.02851
  [hep-th]} \BibitemShut {NoStop}%
\end{thebibliography}%

\end{document}